\documentclass[twocolumn]{aastex62}
\usepackage{amsmath}
\usepackage{amssymb}
\usepackage{empheq}
\usepackage{mathrsfs}
\usepackage[utf8]{inputenc}
\usepackage[T1]{fontenc}			
\usepackage{amssymb}

\newcommand{\appropto}{\mathrel{\vcenter{
  \offinterlineskip\halign{\hfil$##$\cr
    \propto\cr\noalign{\kern2pt}\sim\cr\noalign{\kern-2pt}}}}}
\newcommand{\G}{\mathcal{G}}
\newcommand{\Mdot}{\dot{M}}
\newcommand{\M}{M_{\circ}}

\newcommand{\R}{\mathcal{R}}
\newcommand{\Rc}{R_{\rm{c}}^{\rm{B}}}
\newcommand{\Rch}{R_{\rm{c}}^{\rm{H}}}

\newcommand{\Rb}{R_{\rm{B}}}
\newcommand{\Sigmad}{\Sigma_{\bullet}}
\newcommand{\Rjup}{R_{\rm{Jup}}}
\newcommand{\Mjup}{M_{\rm{Jup}}}

\newcommand{\cs}{c_{\rm{s}}}
\newcommand{\vk}{v_{\rm{K}}}
\newcommand{\Omegak}{\Omega}

\newcommand{\Rh}{R_{\rm{H}}}
\newcommand{\Rt}{R_{\rm{T}}}

\newcommand{\tfric}{t_{\rm{fric}}}

\newcommand{\be}{\begin{equation}}
\newcommand{\ee}{\end{equation}}

\def\lta{\,\raise 0.3 ex\hbox{$ < $}\kern -0.75 em
 \lower 0.7 ex\hbox{$\sim$}\,}
\def\gta{\,\raise 0.3 ex\hbox{$ > $}\kern -0.75 em
 \lower 0.7 ex\hbox{$\sim$}\,}

\begin{document}

%\title{Notes: Slow Gravitational Encounters}

\title{Formation of Giant Planet Satellites}

\author{Konstantin Batygin}
\affiliation{Division of Geological and Planetary Sciences California Institute of Technology, Pasadena, CA 91125, USA}

\author{Alessandro Morbidelli}
\affiliation{Laboratoire Lagrange, Universit\'e C\^ote d'Azur, Observatoire de la C\^ote d'Azur, CNRS, CS 34229, F-06304 Nice, France}

\begin{abstract}
Recent analyses have shown that the concluding stages of giant planet formation are accompanied by the development of large-scale meridional flow of gas inside the planetary Hill sphere. This circulation feeds a circumplanetary disk that viscously expels gaseous material back into the parent nebula, maintaining the system in a quasi-steady state. Here we investigate the formation of natural satellites of Jupiter and Saturn within the framework of this newly outlined picture. We begin by considering the long-term evolution of solid material, and demonstrate that the circumplanetary disk can act as a global dust trap, where $s_{\bullet}\sim0.1-10\,$mm grains achieve a hydrodynamical equilibrium, facilitated by a balance between radial updraft and aerodynamic drag. This process leads to a gradual increase in the system's metallicity, and eventually culminates in the gravitational fragmentation of the outer regions of the solid sub-disk into $\mathcal{R}\sim100\,$km satellitesimals. Subsequently, satellite conglomeration ensues via pairwise collisions, but is terminated when disk-driven orbital migration removes the growing objects from the satellitesimal feeding zone. The resulting satellite formation cycle can repeat multiple times, until it is brought to an end by photo-evaporation of the parent nebula. Numerical simulations of the envisioned formation scenario yield satisfactory agreement between our model and the known properties of the Jovian and Saturnian moons.
\end{abstract}


\section{Introduction} \label{sec:intro}

With the tally of confirmed extrasolar planets now firmly in the thousands \citep{2018ApJS..235...38T}, the feeling of astonishment instigated by the disparity between orbital architectures that comprise the galactic planetary census and that of our own solar system is difficult to resist. Indeed, the widespread detection of planets that complete their orbital revolutions in a matter of days, and have masses on the order of $10-100\,$ppm of their host stars\footnote{For a sunlike star, this mass ratio corresponds to $M\sim3-30\,M_{\oplus}$ planets.}, has inspired a large-scale re-imagination of the dominant physical processes that make up the standard model of planet formation \citep{MorbidelliRaymond2016,JohansenLambrechts2017}. In hindsight, however, the prevalence of such planetary architectures was already foreshadowed by Galilleo's discovery of Jupiter's regular satellites, and Huygens' subsequent discovery of Titan in orbit around Saturn, some four centuries ago. 

Characterized by orbital periods that range from approximately two days (Io) to slightly in excess of two weeks (Callisto and Titan), as well as cumulative masses that add up to about $0.02\%$ of their host planets, both the physical and orbital machinery of giant planet satellites eminently reflect the properties of typical planetary systems found throughout the Galaxy \citep{LaughlinLissauer2015}. Even the intra-system uniformity inherent to the demographics of sub-Jovian extrasolar planets \citep{Weiss2018,Millholland2017} is aptly reproduced in the Galilean ensemble of moons. The ensuing possibility that this similarity may point to a deeper analogy between conglomeration pathways of solar system satellites and short-period exoplanets has not eluded the literature \citep{Kane2013,RonnetJohansen2020}. Nevertheless, it is intriguing to notice that while the pursuit to quantify the formation of extrasolar super-Earths has received considerable attention over the course of the recent decade (see e.g., the recent works of \citealt{Izidoro2019,2019A&A...632A...7L,Bitsch2019,RosenthalMurrayClay2019,2020MNRAS.491.5595P,2020A&A...633A..81K} and the references therein), a complete understanding of the formation of the solar system's giant planet satellites themselves remains incomplete \citep{CanupWard2009,MiguelIda2016,RonnetJohansen2020}. Outlining a new theory for the their conglomeration is the primary purpose of this paper.

\smallskip

Much like the prevailing narrative of the solar system's formation, the theory of satellite formation traces its roots to the nebular hypothesis \citep{Kant1755,Laplace1796}. Both the Galilean moons, as well as Titan, are generically thought to have originated in dissipative disks of gas and dust that encircled Jupiter and Saturn during the first few millions of years of the solar system's lifetime. Within these circumplanetary disks, dust is envisioned to have solidified into satellitesimals through some physical mechanism, and assisted by gravitational and hydrodynamic processes, the satellitesimals  eventually grew into the moons we observe today \citep{1999ARA&A..37..533P}. The devil, however, is in the details, and broadly speaking, current theories of natural satellite formation fall into two categories: the \textit{minimum mass model} and the \textit{gas-starved model}. Let us briefly review the qualitative characteristics of these theories.

\subsection{Existing Models} 
The minimum mass model, first proposed by \citet{LunineStevenson1982}, posits that the cumulative present-day mass of the satellites approximately reflects the primordial budget of solid material that was entrained within the giant planets' circumplanetary disks. Correspondingly, under the assumption of nearly solar metallicity, the minimum mass model entails disks that were only a factor of $\sim50$ less massive than the giant planets themselves, and therefore verge on the gravitational stability limit of nearly Keplerian systems \citep{Safronov1960,Toomre1964}. Generically, this picture is characterized by high disk temperatures and very short satellite conglomeration timescales. 

A somewhat more modern extension of this model was considered by \citet{MosqueiraEstrada2003a,MosqueiraEstrada2003b}, who proposed a circumplanetary nebula that is broken up into a dense inner component, and a more tenuous outer region. Within this model, Galilean satellites are imagined to form from $\sim1{,}000\,$km seeds by accretion of smaller inward-drifting satellitesimals, but only the inner three embryos -- which form in the dense inner disk -- suffer convergent orbital evolution that locks them into the Laplace resonance. Meanwhile, the nebula envisioned by \citet{MosqueiraEstrada2003a,MosqueiraEstrada2003b} is almost perfectly quiescent by construction, such that the steep break in the surface density can persist for the entire lifetime of the system, acting as an effective barrier that halts Callisto's disk-driven orbital decay\footnote{Within the broader framework of satellite-disk interactions, a large surface density gradient leads to a dramatic enhancement in the corotation torque. In turn, this effect preferentially pulls the migrating object into the region of higher density \citep{Masset2006,Paardekooper2018}. Therefore, the surface density jump envisioned by \citet{MosqueiraEstrada2003a,MosqueiraEstrada2003b} is likely to operate as a reversed planet trap, briefly accelerating -- instead of halting -- Callisto's orbital decay.}. 

The ideas outlined by \citet{MosqueiraEstrada2003a,MosqueiraEstrada2003b} were explored in a systematic manner by \citet{MiguelIda2016}. Varying an impressive range of parameters within their simulations, \citet{MiguelIda2016} have convincingly demonstrated that even if one allows for drastic tailoring of the physical state of the system (e.g., ad-hoc modification of the migration timescale as well as solid-to-gas ratio by orders of magnitude), the formation of a satellite system that resembles the real Galilean moons remains exceptionally unlikely within the context of the minimum mass model.

The gas-starved model, put forward by \citet{CanupWard2002} proposes a markedly different scenario. In this picture, the circumplanetary disk is not treated as a closed system, and is assumed to actively interact with its environment, continuously sourcing both gaseous and solid material from the solar nebula. Correspondingly, as solid material is brought into the system, it is envisioned to accrete into large bodies, which -- upon becoming massive enough -- experience long-range inward orbital decay due to satellite-disk interactions. In this manner, the circum-Jovian disk considered by \citet{CanupWard2002} is not required to retain the full mass budget of the Galilean satellites at any one time, and can instead remain in quasi-steady state with relatively low density. 

A key advantage of the gas-starved model is the combination of a comparatively long satellite migration timescale and a sufficiently low disk temperature for effective growth of icy bodies. Because objects that reach the inner edge of the disk are assumed to get engulfed by the planet\footnote{We note that contrary to this assumption, \citet{Sasaki2010} argue that the planetary magnetosphere may effectively truncate the circumplanetary disk, halting the inward migration of satellites at the inner edge.}, the concurrent operation of these two processes (i.e., conglomeration and migration) determines a characteristic steady-state mass scale of the satellites that occupy the disk. Impressively, within the framework of the \citet{CanupWard2002,CanupWard2006} scenario, this quantity evaluates to approximately one ten-thousands of the host planet mass, in agreement with observations.

Despite the numerous successes of the gas-starved model in explaining the basic architecture of the solar system's population of natural satellites, recent progress in theoretical modeling of circumplanetary disk hydrodynamics \citep{Tanigawa2012,Morbidelli2014,Judit2014,Judit2016,Lambrechts2019} has revealed a number of intriguing new challenges pertinent to the formation of giant planet satellites. The most pivotal of the fledging issues concerns the accumulation of sufficient amount of solid material within the disk. In particular, both numerical simulations (e.g., \citealt{Tanigawa2012,Morbidelli2014}) as well as direct observations \citep{Teague2019} show that gas is delivered into the planetary Hill sphere via meridional circulation that is sourced from a region approximately one pressure scale-heigh above the mid-plane of the parent circumstellar nebula. Because of preferential settling of solids to the mid-plane (e.g., \citealt{LambrechtsJohansen2012}), solid material that enters the planetary region is both scarce (implying a sub-solar metallicity of the gas), and small enough (i.e., much smaller than $\sim0.1\,$mm) to remain suspended at large heights within the parent nebula. This picture stands in stark contrast with the scenario of \citet{CanupWard2002,CanupWard2006}, where $\lesssim1\,$m satellitesimals are envisioned to get captured by the circumplanetary disk.

A second problem inspired by the emerging view of giant planet-contiguous hydrodynamics concerns the formation and growth of satellitesimals themselves. That is to say, neither the process by which incoming dust gets converted into satellite building blocks, nor the mechanism through which these solid debris coalesce within strongly sub-Keplerian circumplanetary disks to form the satellites is well understood. Finally, arguments based upon the accretion energetics of the giant planet envelopes as well as considerations of angular momentum transport during the final stages of the planetary growth suggest that primordial magnetospheres of Jupiter and Saturn could have effectively truncated their circumplanetary disks (see for example, \citealt{Batygin2018}). As pointed out by \citet{Sasaki2010}, the resulting cavities within the circumplanetary nebulae could feasibly disrupt the accretion-migration-engulfment cycle envisioned by \citet{CanupWard2006}.

%these models in explaining the general architecture of the giant planet satellite systems, recent progress in theoretical modeling of circumstellar and circumplanetary disk hydrodynamics (cite: Judit, Elena, Michiel, Tanigawa?), as well as planet/satellite-disk interactions (cite: Parderkooper, Bert), has revealed considerable drawbacks that emerge within both of these narratives. In the context of the minimum mass model, ...

%EXPLAIN Note that the co-accretion scenario was actually proposed earlier by A. Harris.

%Despite the successes of these models in explaining the general architecture of the giant planet satellite systems, recent progress in theoretical modeling of circumstellar and circumplanetary disk hydrodynamics (cite: Judit, Elena, Michiel, Tanigawa?), as well as planet/satellite-disk interactions (cite: Parderkooper, Bert), has revealed considerable drawbacks that emerge within both of these narratives. In the context of the minimum mass model, ...

%The complications associated with the gas-starved model are considerably less severe, but are nevertheless significant enough to warrant discussion. 

\subsection{This Work} 
Motivated by the aforementioned developments, in this work, we re-examine the dynamical states of circumplanetary disks during the giant planets' infancy, and propose a new model for the conglomeration of giant planet satellites. We start from first principles, and throughout the manuscript, consistently focus on the characterization of the dominant physical processes, attempting not to prioritize any specific scenario for the evolution of the satellites. Nevertheless, as we demonstrate below, considerations of the basic gravito-hydrodynamic machinery of the proto-satellite disks naturally lends itself to a self-consistent picture for the origins of Jovian and Kronian moons\footnote{The satellite systems of Uranus and Neptune are beyond the scope of our study, as they likely have a distinct origin from the scenario considered herein (see e.g., the recent work of \citealt{2020arXiv200313582I}).}. 

Put succinctly, our model envisions the gradual accumulation of icy dust in a vertically-fed \textit{decretion} disk\footnote{Contrary to accretion disks -- where long-term viscous evolution leads to gradual sinking of nebular material towards the central object -- decretion disks are systems where gas and dust are slowly expelled outwards.} that encircles a newly formed giant planet. Buildup of solid material within the circumplanetary nebula is driven by a hydrodynamical equilibrium, which arises from a balance between viscous outflow of the gas along the disk's mid-plane (that drives dust outward), and sub-Keplerian headwind (that saps the dust of its orbital energy). The cancellation of these two effects allows particles with an appropriate size-range to remain steady within the system. As the cumulative mass of solids within the disk slowly grows, the dust progressively settles towards the mid-plane of the circumplanetary disk under its own gravity. Eventually, gravitational collapse ensues, generating large satellitesimals that are comparable in size to Saturn's small moons. 

\begin{figure*}[tbp]
\centering
\includegraphics[width=\textwidth]{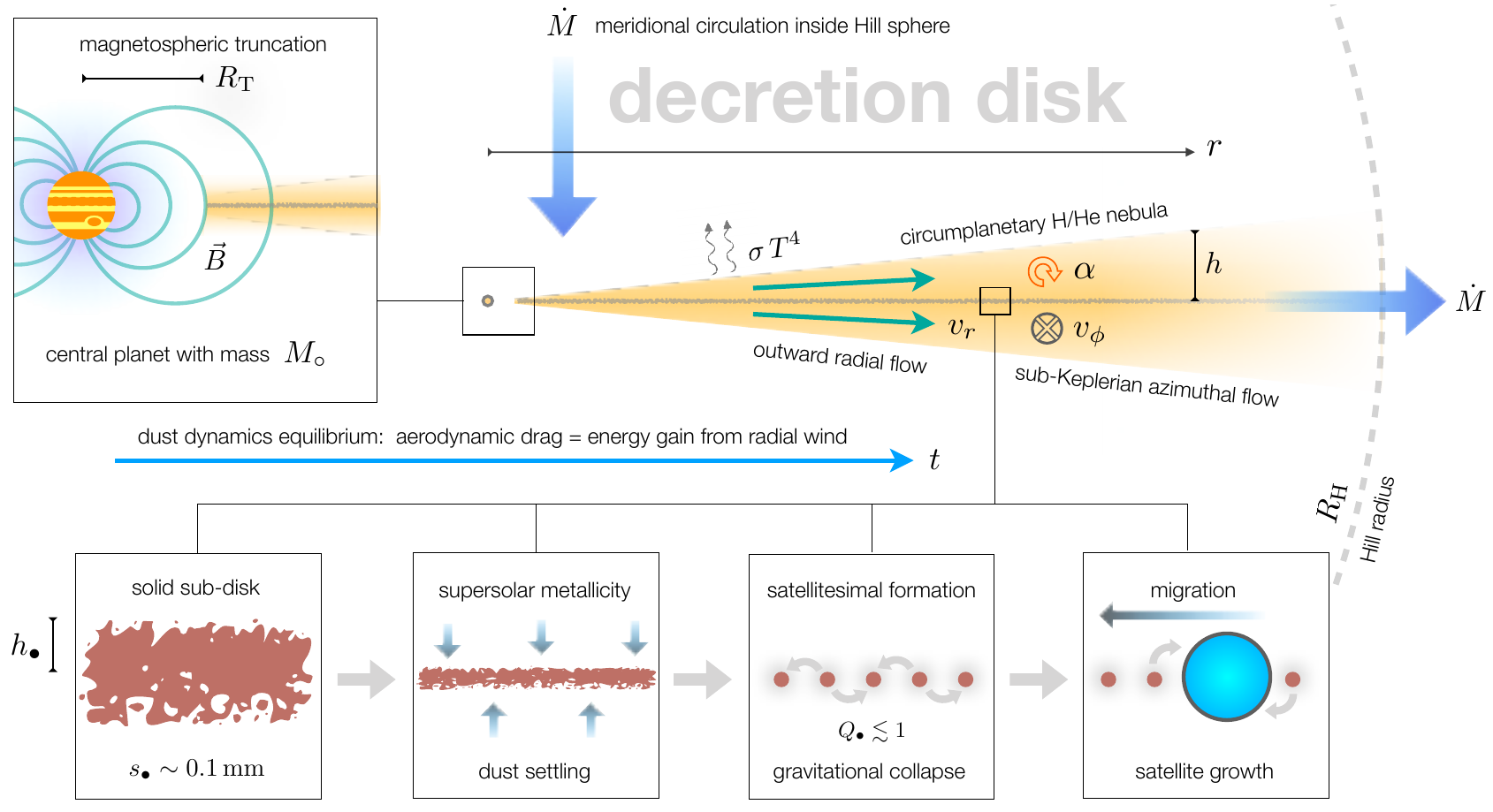}
\caption{A qualitative sketch of our model. A giant planet of mass $\M\sim10^{-3}M_{\odot}$ is assumed to have cleared a gap within its parent nebula, resulting in a steady-state azimuthal/meridional circulation of gas within the planet's Hill sphere. As nebular material overflows the top of the gap (at an altitude of approximately one pressure scale-height above the mid-plane), it free-falls towards the planet -- a process we parameterize by an effective mass-flux $\dot{M}\sim0.1\M/$Myr. Due to conservation of angular momentum, this material spins up, and forms a circumplanetary disk. The inner edge of this disk is truncated by the planetary magnetic field, $B\sim1{,}000\,$G, at a radius $\Rt\sim5\,R_{\rm{Jup}}$. Owing to (magneto-)hydrodynamic turbulence within the circumplanetary disk -- parameterized via the standard $\alpha\sim10^{-4}$ prescription -- the disk spreads viscously and settles into a steady pattern of decretion back into the circumstellar nebula. \\ A balance of viscous heating and radiative losses within the disk determines the system's aspect ratio, $h/r\sim0.1$. The associated pressure support gives rise to sub-Keplerian rotation of the gas, such that the radial and azimuthal components of the circumplanetary flow are $v_r\sim10^{-5}\,\vk>0$ and $v_{\phi}\sim0.99\,\vk<\vk$, respectively. For a critical dust size $s_{\bullet}\sim0.1-10\,$mm, aerodynamic energy loss from the sub-Keplerian headwind exactly cancels the energy gain from the radial wind, trapping the incoming dust within the disk. Through this process, dust accumulates within the system, and disk metallicity, $\mathcal{Z}$, grows in time. This gradual enhancement of the dust-to-gas ratio causes the solid sub-disk to settle ever closer to the mid-plane, and once its scale-height, $h_{\bullet}$, falls below the threshold of gravitational stability ($Q_{\bullet}\lesssim1$), the outer regions of the solid sub-disk fragment into a swarm of $m\sim10^{19}\,$kg satellitesimals. Accretion of satellite embryos proceeds through pairwise collisions, aided by gravitational focusing. Conglomeration is terminated once a satellite grows sufficiently massive to raise appreciable wakes within the circumplanetary disk. At this point, long-range disk-driven migration ensues, ushering the newly formed object towards the magnetospheric cavity.
}
\label{F:Setup} 
\end{figure*}

Mutual collisions among satellitesimals facilitate oligarchic growth, generating satellite embryos. Upon reaching a critical mass -- determined by an approximate correspondence between the timescale for further accretion and the orbital migration time -- newly formed satellites suffer long-range orbital decay, which terminates when the bodies reach the vicinity of the disk's magnetospheric cavity. Owing to continuous aerodynamic damping of the satellitesimal velocity dispersion, the satellite conglomeration process can repeat multiple times, but necessarily comes to a halt after the photoevaporation front within the circumstellar nebula reaches the giant planet orbit. A qualitative sketch of our model is presented in Figure (\ref{F:Setup}).

In the remainder of the paper, we spell out the specifics of our theory. In section \ref{sec:disk}, we outline the model of the circumplanetary disk. The dynamics of dust within the model nebula are discussed in section \ref{sec:dust}. Section \ref{sec:tesimal} presents a calculation of satellitesimal formation. Oligarchic growth of the satellite embryos and orbital migration are considered in section \ref{sec:growth}. Section \ref{sec:sims} presents a series of numerical experiments that quantify the conglomeration of the satellites themselves, as well as the formation of the Laplace resonance. We conclude and discuss the implications of our proposed picture in section \ref{sec:disc}. %Simplicity is emphasized throughout the manuscript, and numerical simulations are only invoked as an aid to confirm/clarify the largely illustrative, analytical framework of our model.

\section{Model Circumplanetary Disk} \label{sec:disk}

The stage for satellite formation is set within the circumplanetary disk, and the construction of a rudimentary model for its structure is the foundational step of our theory. Recent advances in high-resolution numerical hydrodynamics simulations of quasi-Keplerian flow around giant planets \citep{Tanigawa2012,Judit2016} have revealed a nuanced pattern of fluid motion that develops in the vicinity of a massive secondary body when it is embedded within a circumstellar nebula. 

First, as a consequence of gravitational torques exerted by the planet on its local environment, the gas surface density drastically diminishes in the planet's orbital neighborhood, clearing a gap within the circumstellar disk \citep{Crida2006,FungChiang2016}. Within the gap itself, a meridional circulation ensues close to the planet, such that gaseous material rains down towards the planet in a quasi-vertical matter, from an altitude of approximately one disk pressure scale-height\footnote{It is worth noting that recent high-resolution numerical experiments \citep{Lambrechts2019} suggest that for a Jupiter-mass object, such circulation only operates for a sufficiently low mass-accretion rate, which translates to an epoch when the bulk of the planetary mass has already been acquired. Within the context of our model, this means that our envisioned scenario is set towards the last $\sim$Myr of the solar nebula's lifetime.} \citep{Tanigawa2012,Morbidelli2014,Judit2016}. Owing to angular momentum conservation, this material spins up as it free-falls, consolidating into a sub-Keplerian circumplanetary disk. This disk spreads viscously, decreating outwards, such that the constituent gas gets recycled back into the circumstellar nebula (Figure \ref{F:Setup}). 

Importantly, the results of the aforementioned hydrodynamical calculations have shown a remarkable degree of agreement with contemporary observations. In particular, resolved disk gaps -- routinely attributed to dynamical clearing by giant planets -- have become a staple of both sub-mm continuum maps, as well as scattered light images of protoplanetary nebulae (\citealt{2018ApJ...860...27I,Zhang2018} and the references therein). Moreover, the recent detection of the first circumplanetary disk in the PDS\,70 system \citep{2019ApJ...879L..25I} as well as characterization of meridional circulation of gas through observations of $^{12}$CO emission in the HD\,163296 system \citep{Teague2019} lend further credence to the qualitative picture outlined above. 

Swayed by the emergent census of circumplanetary disk simulations and observations, here we adopt an analytic model for a constant $\Mdot$ decretion disk as our starting point. This model was first developed within the context of Be stars by \citet{Lee1991}, and is related to the routinely utilized constant $\Mdot$ accretion disk model \citep{Armitage2010}. However, the differences between the decretion and accretion models are sufficiently subtle that it is worthwhile to sketch out the model's derivation. 

%\paragraph{Surface Density Profile} 
We begin by recalling the continuity equation for the surface density, $\Sigma$, of the circumplanetary disk \citep{1991MNRAS.248..754P}:
\begin{align}
%r\,\frac{\partial\,\Sigma}{\partial\,t} + \frac{\partial}{\partial\,r}(r\,\Sigma\,v_{r}) = \Mdot\, \frac{\delta(r-\Rt) - \delta(r-\Rh)}{2\,\pi},
r\,\frac{\partial\,\Sigma}{\partial\,t} + \frac{\partial}{\partial\,r}(r\,\Sigma\,v_{r}) = \mathcal{S},
\label{eqn:masscontinuity}
\end{align}
where $r$ is the planetocentric distance, and $v_{r}$ is the radial velocity of the fluid. The RHS of the above expression is a $\delta-$function source term that is only non-zero at the inner and outer boundaries of the disk, which we take to be the radius of the magnetospheric cavity, $\Rt$, and the Hill radius, $\Rh$, respectively. In other words, despite the fact that the meridional flow spans a broad range in $r$, here we adopt the simplifying assumption that the vertical flux of material into the disk is localized to $r<\Rt$. Explicitly, the two aforementioned values -- which serve as the confines of our model -- are calculated as follows:
\begin{align}
&\Rt=\bigg(\frac{\zeta^7}{2\mu_0}\frac{\mu^4}{\G\,\M\,\Mdot^2} \bigg)^{1/7} &\Rh=a\bigg(\frac{\M}{3M_{\odot}}\bigg)^{1/3},
\label{eqn:RtRh}
\end{align}
where $\zeta$ is a dimensionless constant of order unity, $\mu_0$ is the permeability of free space, while $a$, $\M$ and $\mu$ are the planet's heliocentric semi-major axis, mass, and magnetic moment, respectively \citep{MohantyShu2008}. For definitiveness, here we adopt Jovian parameters, noting that the Hill radii of Jupiter and Saturn are approximately one-third and one-half of an AU respectively, while for system parameters relevant to the final stages of runaway accretion, the disk's magnetospheric truncation radius evaluates to $\Rt\sim5\,\Rjup$ -- a value marginally smaller than Io's present-day semi-major axis \citep{Batygin2018}.

In steady state ($\partial/\partial\,t\rightarrow 0$), the first term of equation (\ref{eqn:masscontinuity}) vanishes, such that in the region of interest ($\Rt<r<\Rh$), the solution is simply given by 
\begin{align}
r\,\Sigma\,v_{r}=\frac{\Mdot}{2\,\pi}=\mathrm{const}.
\label{eqn:mdot}
\end{align}
This expression establishes a connection between the surface density and radial fluid velocity at all relevant radii, parameterized by the rate at which mass flows through the system, which we set to $\Mdot=0.1\,\Mjup/$Myr. We emphasize that this value of $\Mdot$ is low compared with the characteristic mass-accretion rate of nebular material onto T-Tauri stars (which is closer to $\Mdot_{\star}\sim10\,\Mjup/$Myr), and is appropriate only for the concluding stage of the circumstellar disk's evolution, when our model is envisioned to operate.

The continuity equation for angular momentum within the disk is written as follows \citep{1974MNRAS.168..603L}:
\begin{align}
r\frac{\partial\,(r^2\,\Sigma\,\Omegak)}{\partial\,t}+\frac{\partial\,(r^3\,\Sigma\,v_r\,\Omegak)}{\partial\,r}=\frac{\partial}{\partial\,r}\bigg(r^3\,\nu\,\Sigma\,\frac{\partial\,\Omegak}{\partial\,r} \bigg),
\label{eqn:amcontinuity}
\end{align}
where $\nu$ is the viscosity and $\Omegak=\sqrt{\G\,\M/r^3}$ is the Keplerian orbital frequency (i.e., mean motion) of the disk material. As before, the steady state assumption eliminates the leading term of equation (\ref{eqn:amcontinuity}), and upon substituting the definition of $\Mdot$ and taking the derivative, we re-write the momentum continuity equation as follows \citep{Lee1991}:
\begin{align}
\frac{\partial\,(r^2\,\mathcal{W})}{\partial\,r}+\frac{\Mdot}{4\,\pi}\sqrt{\frac{\G\,\M}{r}}=0,
\label{eqn:amcontinuityW}
\end{align}
where $\mathcal{W}=-r\, \nu\, \Sigma\, \partial \Omegak/ \partial r$ is identified as the vertically integrated viscous stress tensor.

\begin{figure}[tbp]
\centering
\includegraphics[width=\columnwidth]{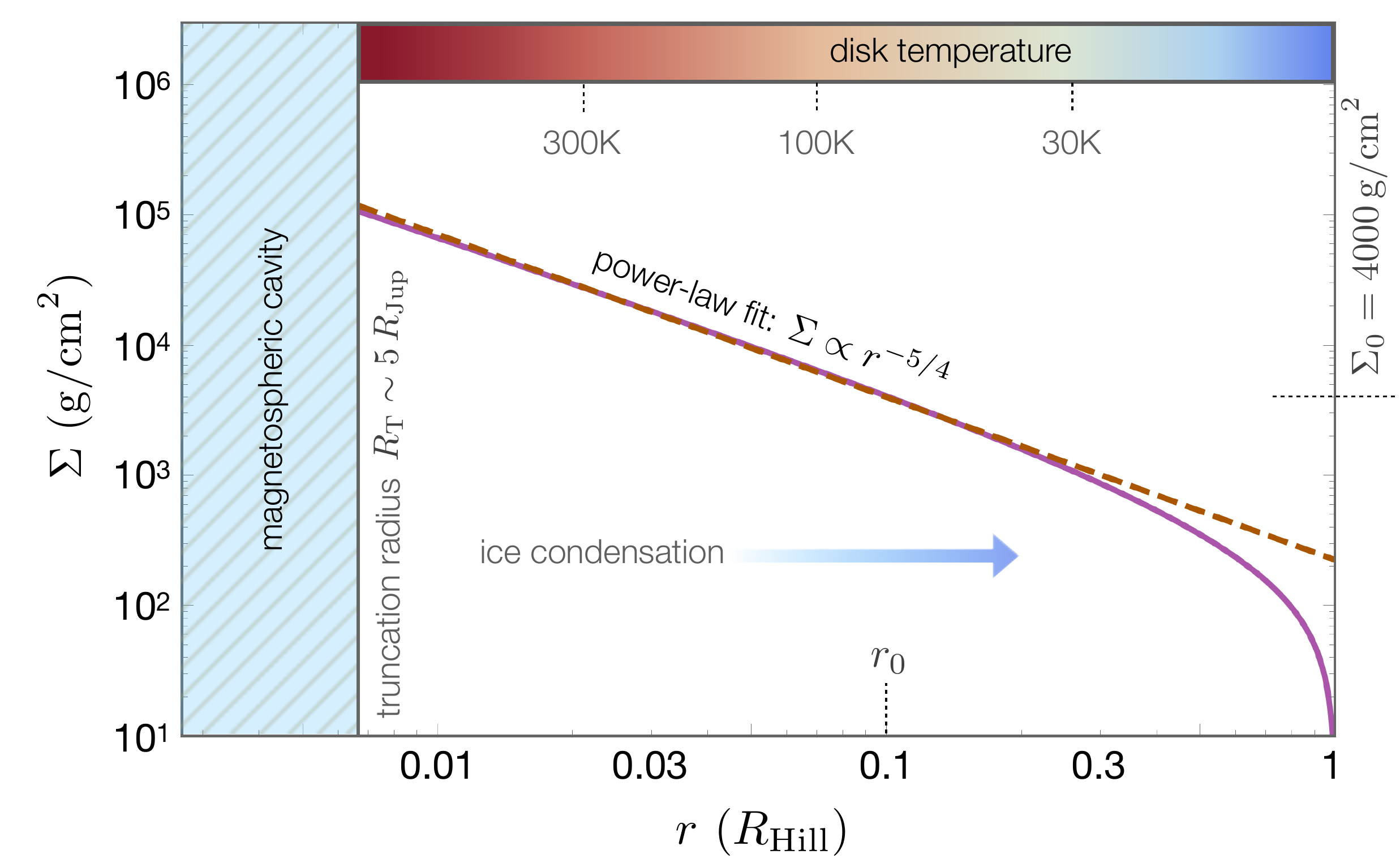}
\caption{Surface density profile of our model circumplanetary disk. The exact solution (equation \ref{eqn:sigma}) -- corresponding to a constant-$\dot{M}$ decretion $\alpha-$model is shown with a solid purple line. An index $-5/4$ power-law fit to this solution (equation \ref{eqn:powerlaw}) is depicted with a dashed dark orange line. The reference surface density $\Sigma_0=4{,}000\,$g/cm$^2$ and reference radius $r_0=0.1\Rh$ are marked with thin dotted lines. The (assumed vertically isothermal) temperature profile of the disk is represented at the top of the figure with a color bar. Importantly, within the context of our model, temperatures are sufficiently low for ice condensation to ensue beyond $r\gtrsim r_0$. The inner edge of the disk is determined by the size of the magetospheric cavity (equation \ref{eqn:RtRh}), which we take to be $\Rt\sim5\,R_{\rm{Jup}}$.}
\label{F:Sigma} 
\end{figure}

Equation (\ref{eqn:amcontinuityW}) is readily solved for $\mathcal{W}$ as a function of $r$, upon specification of a single boundary condition. To this end, we assume that the viscous torque vanishes at the outer edge of the disk, such that $\mathcal{W}=0$ at $r=\Rh$. We then have:
\begin{align}
%\mathcal{W}=-r\,\nu\,\Sigma\,\frac{\partial\,\Omegak}{\partial\,r}=\frac{\Mdot}{2\pi}\,\Omegak\,\bigg(\sqrt{\frac{\Rh}{r}}-1 \bigg).
\mathcal{W}=\frac{\Mdot}{2\pi}\,\Omegak\,\bigg(\sqrt{\frac{\Rh}{r}}-1 \bigg).
\label{eqn:Wsolution}
\end{align}
Noting that $\partial \Omegak/\partial r=-3\Omega/(2\,r)$ and cancelling the dependence on $\Omegak$ in the above expression, we obtain the surface density profile of the disk (Figure \ref{F:Sigma}):
\begin{align}
\boxed{\Sigma=\frac{\Mdot}{3\,\pi\,\nu}\,\bigg(\sqrt{\frac{\Rh}{r}}-1 \bigg).}
\label{eqn:sigma}
\end{align}

We note that herein lies an important difference between constant $\Mdot$ accretion and decretion disks. For an accretion disk where $\Mdot$ is negative, the relation
\begin{align}
\frac{\Mdot}{2\,\pi\,r}=-\frac{3}{\sqrt{r}}\frac{\partial \,(\nu\,\Sigma\,\sqrt{r})}{\partial \,r}
\label{eqn:morbybeer}
\end{align}
is satisfied by $\nu\,\Sigma=\rm{const}.$ This, however, cannot hold true in principle for a decretion disk. That is, since radial motion of the disk material is facilitated by viscous spreading, equation (\ref{eqn:morbybeer}) necessitates that $\nu\,\Sigma$ must be a function that decays more steeply in radius than $\sqrt{r}$, for $\Mdot$ (and by extension, $v_r$) to be positive.

%Note that the functional form of equation (\ref{eqn:sigma}) is inversely proportional to the viscosity. 

In order to complete the specification of the problem, we must define the functional form of the viscosity, and to do so we adopt the standard \citet{1973A&A....24..337S} $\alpha$ prescription, setting 
\begin{align}
\nu=\alpha\,\cs\,h = \alpha\, \Omegak\,h^2,
\label{eqn:alpha}
\end{align}
where $\cs=\sqrt{k_{\rm{b}}\,T/\mu}=h/\Omegak$ is the isothermal speed of sound and $h$ is the pressure scale-height. Of course, the value of $\alpha$ itself is highly uncertain. Nevertheless, we note that recent results from the DSHARP collaboration \citep{Dullemond2018} report lower limits on turbulent viscosity within circumstellar disks that translate to $\alpha\sim10^{-4}$. Following this work, here we set $\alpha=10^{-4}$, but note that the surface density profile itself only depends on the ratio of $\Mdot/\alpha$, implying considerable degeneracy between two poorly determined quantities. 

Assuming that the circumplanetary disk is ``active,'' the disk temperature is determined by an energy balance between viscous heat generation within the nebula, $Q_+=(9/8)\,2\,\pi\,r\,\nu\,\Sigma\,\Omegak^2$, and black-body radiative losses from its surface, which for an optically thin disk have the simple form $Q_-=4\,\pi\,r\,\sigma\,T^4$ \citep{Armitage2010}. In turn, recalling the proportionality between temperature and the speed of sound, this equilibrium determines the geometrical aspect ratio $h/r=\cs/\vk$ of the disk:
\begin{align}
%\frac{h}{r}=\frac{\cs}{\vk}=\bigg(\frac{3}{8\,\pi}\bigg)^{1/8}\sqrt{\frac{k_{\rm{b}}}{\G\,\M\,\mu}\left(\frac{\G\,\M\,\Mdot\,r}{\sigma} \left(\sqrt{\frac{\Rh}{r}}-1 \right) \right)^{1/4}}
\boxed{\frac{h}{r}=\sqrt{\frac{k_{\rm{b}}}{\G\,\M\,\mu}\left(\frac{3\,\G\,\M\Mdot\,r}{8\,\pi\,\sigma} \left(\sqrt{\frac{\Rh}{r}}-1 \right) \right)^{1/4}}.}
\label{eqn:hr}
\end{align}
We admit that our assumption of an optically thin nebula is a simplifying one, and caution that it can only be justified if the system's budget of micron-sized dust is low. However, because $\mu$m dust is in general very tightly coupled to the gas, it is unlikely that it can ever accumulate in a steady-state decretion disk, implying that our optically thin and isothermal assumption may be defensible.

\begin{figure}[tbp]
\centering
\includegraphics[width=\columnwidth]{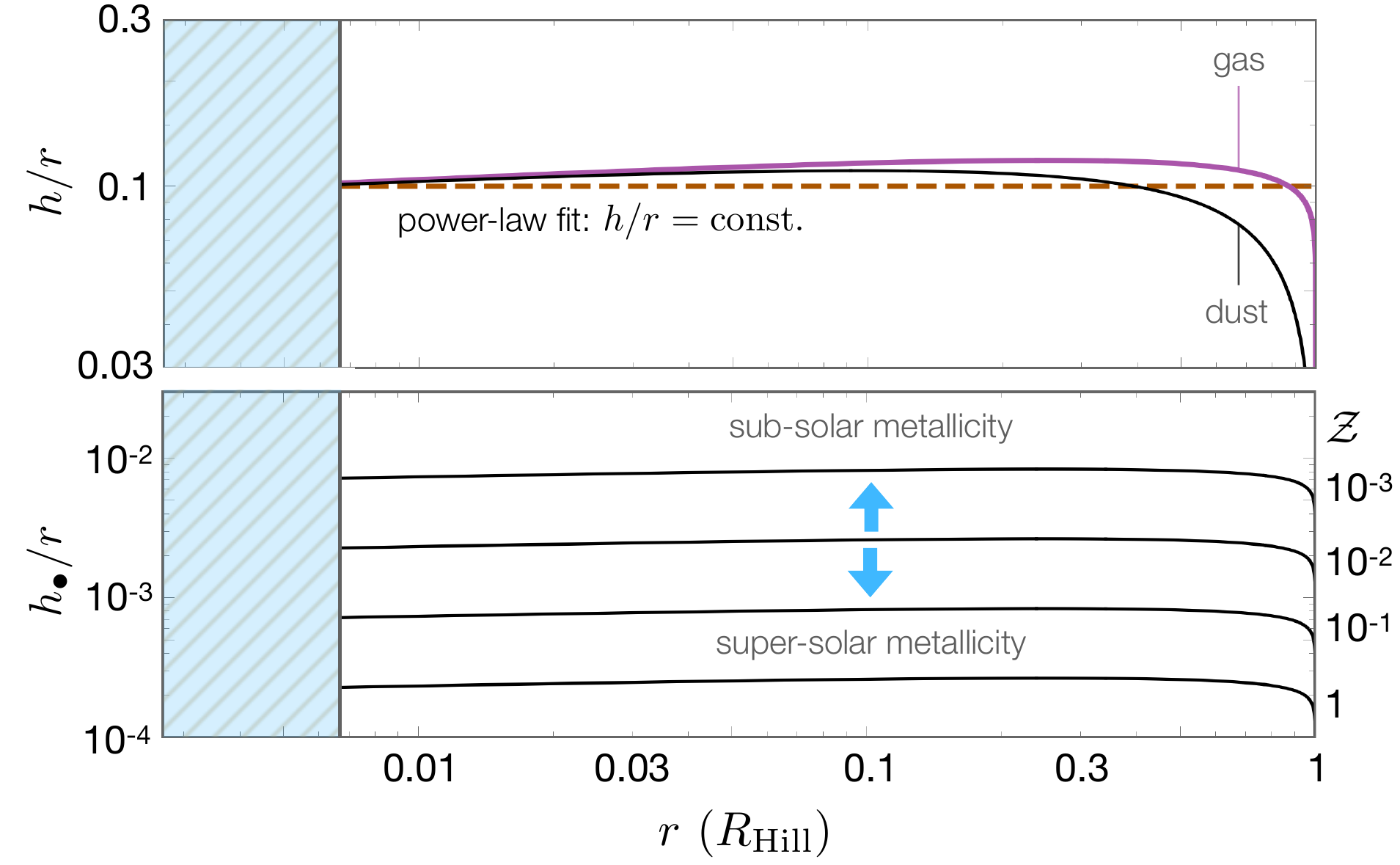}
\caption{Aspect ratio of the circumplaentary nebula. Within the framework of our model, the vertical thickness of the gaseous disk merely reflects its temperature structure (via $h/r=c_{\rm{s}}/\vk\propto\sqrt{T}$). Because viscous energy dissipation dominates over planetary irradiation (equation \ref{eqn:L}), the temperature profile itself is determined by equating turbulent heating within the disk to black-body radiative losses at its surface. The top panel shows the exact solution for the gaseous component of the nebula (equation \ref{eqn:hr}) with a solid purple line. For our purposes, it suffices to ignore minor variations in $h/r$ with orbital radius, and envision the disk as having a constant aspect ratio. Our adopted value of $h/r=0.1$ is shown with a dashed dark orange line on the top panel. The aspect ratio of the dust layer -- computed in the massless tracer-particle limit with $\rm{Sc}$ of unity (equation \ref{eqn:hs}) -- is shown on the top panel with a black curve. The bottom panel shows the aspect ratio of the solid sub-disk, $h_{\bullet}/r$, for a variety of disk metallicities, accounting for energetic suppression of the dust disk's thickness (equation \ref{eqn:hrenergy}). Note that once energetic limitation of turbulent stirring of the dust layer is taken into consideration, $h_{\bullet}/r\ll h/r$. Moreover, it is worthwhile to note that $h_{\bullet}/r\propto1/\sqrt{\mathcal{Z}}$, and that for $\mathcal{Z}\gtrsim0.1$, $h_{\bullet}/r\lesssim10^{-3}$.}
\label{F:hr} 
\end{figure}

Quantitatively, the above expression evaluates to $h/r\sim0.1$ throughout the circumplanetary nebula, implying a relatively thick, almost un-flared disk structure (Figure \ref{F:hr}). We further note that the aspect ratio only exhibits a very weak dependence on $\Mdot$, and no explicit dependence on $\alpha$, insinuating a pronounced lack of sensitivity to poorly constrained parameters. With the expression for the disk scale-height specified, the mid-plane gas density can be calculated in the usual manner: $\rho=\Sigma/(\sqrt{2\,\pi}\,h$).

Of course, the underlying assumptions of equation (\ref{eqn:hr}) are only sensible if viscous heating dominates over planetary irradiation in the region of interest. Quantitatively, this physical regime is appropriate if the ratio between radiative and viscous heating
\begin{align}
\mathcal{L}=\frac{16\,\sigma\,T_{\circ}^4\,R_{\circ}^3}{9\,\G\,\M\,\Mdot\,(\sqrt{\Rh/r}-1)}
\label{eqn:L}
\end{align}
is significantly smaller than unity. Indeed, for planetary parameters of $T_{\circ}=1{,}000\,$K and $R_{\circ}=2\,\Rjup$, $\mathcal{L}<1$ for $r\lesssim\Rh/2$, albeit only by a factor of a few. This means that even though irradiation from the central planet does not dominate the disk's thermal energy balance, it can contribute a notable correction (e.g., minor flaring) to the aspect ratio profile (\ref{eqn:hr}), especially in the outer regions of the circumplanetary nebula. Here, however, we neglect this technicality to keep the model as simple as possible.

While equations (\ref{eqn:sigma}) and (\ref{eqn:hr}) provide exact solutions for the surface density and aspect ratio profiles of the model nebula, much of the literature on astrophysical disks is built around consideration of power-law models, and it is illustrative to make the connection between this simplified description, and the more self-consistent picture outlined above. Thus, adopting $r_0=0.1\,\Rh$ as a reference radius, we find that our model can be crudely represented by the fit: 
\begin{align}
&\Sigma\approx\Sigma_0\,\bigg(\frac{r_0}{r} \bigg)^{5/4} &\bigg(\frac{h}{r} \bigg)\approx0.1,
\label{eqn:powerlaw}
\end{align}
where the reference surface density at $r=r_0$ is $\Sigma_0\approx4{,}000\,$g/cm$^2$. For comparison, we note that the above expression corresponds to a proto-satellite nebula that is slightly steeper than a \citet{Mestel1963} type $\Sigma\propto r^{-1}$ disk, while being marginally shallower than Hayashi-type minimum mass solar nebula \citep{Weidenschilling1977MMSN,Hayashi1981}. These rudimentary profiles are shown alongside equations (\ref{eqn:sigma}) and (\ref{eqn:hr}) in Figures (\ref{F:Sigma}) and (\ref{F:hr}).

A subtle, but important consequence of radial pressure-support, $\partial\,P/\partial\,r$, within the circumplanetary disk is the sub-Keplerian rotation of the gas. A conventional way to parameterize the degree to which the gas azimuthal velocity, $v_\phi$, lags the keplerian velocity, $\vk$, is to introduce the factor (e.g., \citealt{Armitage2010}):
\begin{align}
\eta=-\frac{1}{2}\frac{r^2}{\G\,\M\,\rho}\frac{\partial\,P}{\partial\,r}=\frac{13}{8}\bigg(\frac{h}{r}\bigg)^2\sim\mathcal{O}\big(10^{-2}\big),
\label{eqn:eta}
\end{align}
where the numerical factor on the RHS corresponds to the power-law surface density profile (\ref{eqn:powerlaw}). Accordingly, both the azimuthal and radial velocity of the fluid within the circumplanetary nebula are now defined, and have the form:
\begin{empheq}[box=\fbox]{align}
&v_\phi=\vk\,\sqrt{1-2\,\eta}\approx\vk\,\Bigg(1- \frac{13}{8}\bigg(\frac{h}{r}\bigg)^2\Bigg) < \vk \nonumber \\
&v_r=\frac{\Mdot}{2\,\pi\,r\,\Sigma}\approx\frac{\Mdot}{2\,\pi\,r\,\Sigma_0}\Bigg( \frac{r}{r_0} \Bigg)^{5/4}>0.
\label{eqn:vphivr}
\end{empheq}
With the specification of the model circumplanetary disk complete, let us now examine the evolution of solid dust embedded within this nebula.

\section{Dust Dynamics} \label{sec:dust}

A rudimentary precondition that must be satisfied within the context of giant planet satellite formation theory is the accumulation of sufficient amount of solid material within the circumplanetary disk. As already mentioned in the introduction, however, this basic issue of accumulating the necessary high-metallicity mass budget to form the satellites ($M_{\bullet}/\M\gtrsim2\times10^{-4}$, where $M_{\bullet}$ refers to the total mass of solids within the disk) poses a formidable problem \citep{RonnetJohansen2020}. In fact, the difficulty in capturing icy and rocky matter from the circumstellar nebula is two fold. 

On one hand, the majority of small grains that drift towards the planet's semi-major axis by way of aerodynamic drag get shielded away from the planet's orbital neighborhood due to a local pressure maximum that develops in the planet's vicinity for $\M\gtrsim30\,M_{\oplus}$ \citep{LambrechtsJohansen2014}. More specifically, high-resolution numerical simulations of \citet{2018ApJ...854..153W,2019AJ....158...55H} suggest that only grains smaller than $s\lesssim 0.1\,$mm are sufficiently well coupled to the gas to remain at a high enough altitude in the circumstellar disk to bypass the mid-plane pressure barrier, and enter into the planetary Hill sphere together with the meridional circulation. Even so, in light of their near-perfect coupling to the gas, it is a-priori unclear how such small grains can get sequestered in the circumplanetary disk, instead of getting expelled back into the circumstellar nebula, together with the decretionary flow.

On the other hand, direct injection of planetesimals into the circumplanetary disk appears problematic from an energetic point of view. That is, calculations of \citet{2009euro.book...27E} and \citet{2010SSRv..153..431M} suggest that planetesimals that are successfully captured around the planet inevitably experience large-scale ablation (see also the recent work of \citealt{RonnetJohansen2020}). Accordingly, small grains generated from the process of planetesimal evaporation likely suffer the same fate as grains injected by the meridional circulation. Meanwhile, larger fragments -- even if extant -- are likely to rapidly spiral onto the planet, due to interaction with a strong headwind generated by the appreciably sub-Keplerian flow of the circumplanetary gas \citep{Weidenschilling1977AERO}. 

In this section, we outline how this problem is naturally circumvented within the framework of our model. We begin by writing down the well-studied equations of motion for a solid particle of radius $s_{\bullet}$ in orbit of the central planet, that experiences aerodynamic drag in the Epstein regime, arising from the marginally sub-Keplerian flow of circumplanetary gas \citep{TakeuchiLin2002,LambrechtsJohansen2012}. For simplicity, we ignore the back-reaction of dust upon the gaseous nebular fluid for the time being, but return to this issue below (and quantify it in appendix \ref{appndA}). Under the assumption of a nearly-Keplerian, circular orbit, the azimuthal equation of motion reads:
\begin{align}
\frac{1}{r}\frac{d\,(r\,v_{\phi\,\bullet})}{d\,t}\approx\frac{v_{r\,\bullet}\,\vk}{2\,r}=-\frac{v_{\phi\,\bullet}-v_{\phi}}{\tfric},
\label{eqn:azimuth}
\end{align}
where $\tfric$ is the frictional timescale relevant for the Epstein regime of drag:
\begin{align}
\tfric=\sqrt{\frac{\pi}{8}}\frac{\rho_{\bullet}}{\rho}\frac{s_{\bullet}}{\cs}.
\label{eqn:tfric}
\end{align}
Note that $\tfric$ is approximately the isothermal sound crossing time across particle radius, weighted by the solid-to-gas material density ratio.

A key feature of equation (\ref{eqn:azimuth}) is that the radial velocity of the particle vanishes in the limit where the particle azimuthal velocity matches that of the gas. To examine if such a balance is possible, let us consider the radial equation of motion:
\begin{align}
\frac{d\,v_{r\,\bullet}}{d\,t}=\frac{v_{\phi\,\bullet}^2}{r}-r\,\Omegak^2-\frac{v_{r\,\bullet}-v_r}{\tfric}
\label{eqn:radial}
\end{align}
The inertial term on the LHS can be set to zero, and following \citet{TakeuchiLin2002}, we assume that both $v_{\phi}$ and $v_{\phi\,\bullet}$ are close to $\vk$. Then, writing $r \,\Omega^2$ as $\vk\,(v_{\phi}+\eta\, \vk)/r$, and retaining only leading order terms, we have:
\begin{align}
&2\,\eta\,\frac{\vk^2}{r}=2\,\vk\,\bigg(\frac{v_{\phi\,\bullet}-v_{\phi}}{r}\bigg)-\frac{v_{r\,\bullet}-v_r}{\tfric} \nonumber \\
&=-\frac{\vk^2}{r}\bigg( \frac{v_{r\,\bullet}\,\tfric}{r}\bigg)-\frac{v_{r\,\bullet}-v_r}{\tfric},
\label{eqn:radial2}
\end{align}
where we have employed equation (\ref{eqn:azimuth}) to arrive at the second line of the expression. 

Setting $v_{r\,\bullet}=0$, we trivially obtain an equilibrium solution for a particle's equilibrium frictional timescale, $\tfric^{(\rm{eq})}=r\,v_r/(2\,\eta\,\vk)$. It is further convenient to express this this quantity in terms of the dimensionless frictional time i.e., the Stokes number $\tau^{(\rm{eq})}=\tfric^{(\rm{eq})}\,\Omegak$:
\begin{align}
\boxed{\tau^{(\rm{eq})}=\frac{v_{r}}{2\,\eta\,\vk}\big(1+\mathcal{Z}_{\rm{mid}}\big) =\frac{\Mdot\,\big(1+\mathcal{Z}_{\rm{mid}}\big)}{4\,\pi\,\eta\,\Sigma\,\sqrt{\G\,\M\,r}},}
\label{eqn:equilibrium}
\end{align}
where the correction factor due to the mid-plane metallicity\footnote{The relationship between $\mathcal{Z}_{\rm{mid}}$ and $\mathcal{Z}$ is given by equation (\ref{eqn:Zmid}).}, $\big(1+\mathcal{Z}_{\rm{mid}}\big)$, originates from a marginally more detailed analysis that accounts for the back-reaction of dust upon gas (see Appendix \ref{appndA}). For our adopted disk parameters and global metallicity that falls into the $\mathcal{Z}\sim0.01-0.3$ range, the above expression evaluates to $\tau^{(\rm{eq})}\sim \mathcal{O} \big(10^{-5}-10^{-3}\big)$.

The existence of a stationary $v_{r\,\bullet}=0$ solution to equation (\ref{eqn:radial2}) implies that as solid material enters the circumplanetary nebula either through the meridional circulation or via ablation of planetesimals, dust with a physical radius that satisfies equation (\ref{eqn:equilibrium}) will get trapped in the disk. It is important to understand that qualitatively, this equilibrium stems from a balance between loss of particle angular momentum due to aerodynamic drag and gain of angular momentum due to coupling with a radial outflow of the gas. Consequently, this hydrodynamically facilitated process of dust accumulation can only function in a decretion disk. Indeed, the same process does not operate within circumstellar accretion disks. This discrepancy brings to light an important distinction between formation of Super-Earth type extrasolar planets and solar system satellites: despite having similar characteristics in terms of orbital periods and normalized masses, it is likely that their conglomeration histories are keenly distinct.

Examining the functional form of expression (\ref{eqn:equilibrium}), we note that because $\Sigma$ is proportional to $\Mdot$, the equilibrium Stokes number is independent of the assumed mass-accretion rate. Moreover, from equations (\ref{eqn:sigma}) and (\ref{eqn:powerlaw}), it is easy to see that $\tau^{(\rm{eq})}\appropto \alpha\,r^{3/4}$ and thus varies by less than an order of magnitude over the radius range of interest (i.e., $r\sim0.1-0.3\,\Rh$). Instead, this quantity is largely controlled by the assumed value of the viscosity parameter and the mid-plane metallicity (which is envisioned to slowly increase in time). 

%We further note that the equilibrium value of the particle radius can vary between a fraction of a millimeter and at most a centimeter depending on the assumed value of $\alpha$.

\begin{figure}[tbp]
\centering
\includegraphics[width=\columnwidth]{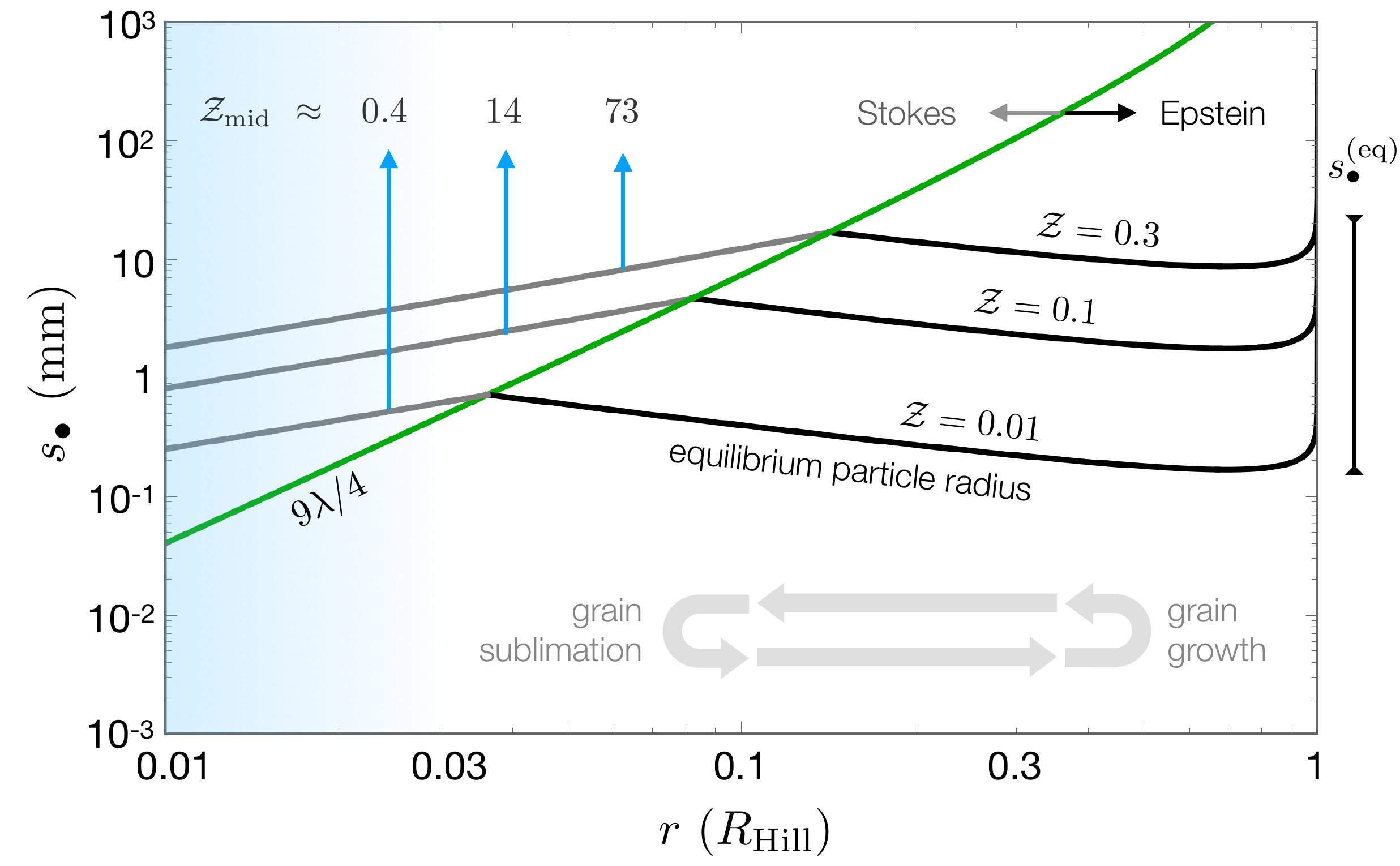}
\caption{Equilibrium dust grain radius, $s_{\bullet}^{(\rm{eq})}$, as a function of planetocentric distance. Dust grains with radii bounded by the denoted range (i.e. $s_{\bullet}\sim0.1-10\,$mm; equation \ref{eqn:equilibrium}) will remain trapped within the circumplanetary disk, thanks to a balance between aerodynamic drag and radial updraft. Equilibrium curves corresponding to global disk metallicities of $\mathcal{Z}=0.01,0.1,$ and $0.3$ (which translate to mid-plane metallicities of $\mathcal{Z}_{\rm{mid}}=0.4,14,$ and $74$ respectively; see equation \ref{eqn:Zmid}) are shown. The green line depicted on the figure marks the transition between Epstein and Stokes regimes of drag: $(s_{\bullet})_{\rm{t}}=9\lambda/4$, where $\lambda$ denotes the mean free path of gas molecules. The derived dust equilibrium is stable for constant particle size in the Epstein regime of drag, but not the Stokes regime. Although not directly modeled, it is likely that the dust growth/sublimation cycle can play an important auxiliary role in modulating $s_{\bullet}$. In particular, one can envision that because $s_{\bullet}^{(\rm{eq})}$ is a decreasing function of $r$ in the Epstein regime, particle growth in the outer disk can cause orbital decay. The reverse effect ensues at small orbital radii, where sublimation of icy grains interior to the ice-line causes particle size to fall below the equilibrium value, expelling solid material outward, where grain growth can ensue once again, thereby maintaining $s_{\bullet}\sim s_{\bullet}^{(\rm{eq})}$ on average.}
\label{F:seq} 
\end{figure}

For our fiducial value of $\alpha=10^{-4}$ and global metallicity\footnote{As we demonstrate below, $\mathcal{Z}=0.01$ translates to $\mathcal{Z}_{\rm{mid}}\lesssim1$, meaning that the correction factor $(1+\mathcal{Z}_{\rm{mid}})$ in equation (\ref{eqn:equilibrium}) is unimportant.} of $\mathcal{Z}=0.01$, the typical equilibrium Stokes number is of order $\tau^{(\rm{eq})}\sim 10^{-5}$ for the relevant disk radii. In terms of physical particle radius (for $\rho_{\bullet}=1\,$g/cc) this value translates\footnote{Because $\tau^{(\rm{eq})}\propto \alpha$, higher values of the Shakura--Sunyaev viscosity parameter would yield proportionally larger $s_{\bullet}^{(\rm{eq})}$. However, given that generically, $\alpha\lesssim0.01$, the equilibrium dust radius for $\mathcal{Z}=0.01$ is unlikely to exceed $\sim$ a few $\rm{cm}$.} to $s_{\bullet}^{(\rm{eq})}\sim\rm{few}\,\times\,10^{-2}\,$cm. More precisely, $s_{\bullet}^{(\rm{eq})}$ is shown as a function of $r$ in Figure (\ref{F:seq}). In addition to a line corresponding to $\mathcal{Z}=0.01$, Figure (\ref{F:seq}) also depicts curves corresponding to super-solar metallicities of $\mathcal{Z}=0.1$ and $\mathcal{Z}=0.3$. The determination that $s_{\bullet}^{(\rm{eq})}$ is always much larger than a micron suggests that the dust-to-gas ratio of the circumplanetary disk can increase dramatically without contributing an associated enhancement to opacity. Thus, the envisioned picture is consistent with our assumption of an optically-thin disk. 

The assumption that interactions between solid particles and gas lie in the Epstein regime is only justified as long as $s_{\bullet}^{(\rm{eq})}\lesssim9\,\lambda/4$ (where $\lambda=1/n\,\sigma$ is the mean free path of gas molecules). From Figure (\ref{F:seq}), it is clear that the equilibrium grain radius given by equations (\ref{eqn:tfric}) and (\ref{eqn:equilibrium}) is smaller than $\lambda$ throughout most of the disk, but not everywhere. This begs the question of what happens to solid material where the Epstein criterion is not satisfied.

In a parameter regime where $s_{\bullet}^{(\rm{eq})}\gtrsim9\,\lambda/4$, we must consider the Stokes regime of aerodynamic drag. For low Reynolds number flow (specifically, $\rm{Re}<1$), the corresponding aerodynamic drag force is given by \citep{Weidenschilling1977AERO}:
\begin{align}
&\mathcal{F}_{\rm{D}}=\frac{12\,\pi}{\rm{Re}}\,s_{\bullet}^2\,\rho\,v_{\rm{rel}}^2 &\mathrm{Re}=\frac{\sqrt{2\,\pi}\, s_{\bullet}\,v_{\rm{rel}}}{\lambda\,\cs}.
\label{eqn:Stokesdrag}
\end{align}
Conveniently, the corresponding frictional timescale -- which replaces the expression given in equation (\ref{eqn:tfric}) when $s_{\bullet}\geqslant9\,\lambda/4$ -- is independent of $v_{\rm{rel}}$, and has the form:
\begin{align}
\tfric=\frac{m_{\bullet}\,v_{\rm{rel}}}{\mathcal{F}_{\rm{D}}}=\frac{\sqrt{2\,\pi}}{9\,\lambda}\frac{s_{\bullet}^2\,\rho_{\bullet}}{\cs\,\rho}.
\label{eqn:Stokestfric}
\end{align}

In Figure (\ref{F:seq}), segments of equilibrium particle size curves that correspond to Epstein and Stokes drag are shown as black and gray lines, respectively. Notably, the two regimes join across an equilibrium grain radius of $s_{\bullet}^{(\rm{eq})}=9\,\lambda/4$, where equations (\ref{eqn:tfric}) and (\ref{eqn:Stokestfric}) are equivalent. It is further worth noting the change in sign of the derivative of $s_{\bullet}^{(\rm{eq})}$ with respect to $r$ across this transition: in the Epstein regime, the equilibrium radius is a decreasing function of the planetocentric radius, whereas the opposite is true in the Stokes regime. As we discuss below, this switch has important implications for the stability of the derived equilibrium.

Of course, this work is by no means the first to propose a dust-trapping mechanism within quasi-Keplerian astrophysical disks. Rather, pressure maxima associated with long-lived vortices, boundaries between turbulent and laminar regions of the system, zonal flows, etc., have been widely discussed as potential sites for localized enhancements of the nebular dust-to-gas ratio (see \citealt{2006A&A...446L..13V,2009ApJ...704L..75J,2009A&A...497..869L,2012A&A...545A..81P,2018ApJ...866..142D} and the references therein). What differentiates the process outlined above from these ideas, however, is the fact that it stems from a balance between two large-scale features of circumplanteary disk circulation, and is therefore global in nature.

If left unperturbed in a quiescent environment, the dust accumulating within the circumplanetary disk would inevitably sink onto the mid-plane. The characteristic timescale on which this occurs is easily obtained from the Epstein drag equation \citep{Armitage2010}, and is simply $\mathcal{T}_{\rm{settle}} = 1/(\tau^{(\rm{eq})}\,\Omegak)$. Disk turbulence, on the other hand, opposes dust settling by stochastically enhancing its vertical velocity dispersion. Given that the same turbulent eddies that drive vertical stirring also facilitate the viscous evolution of the disk, the diffusion coefficient associated with turbulent stirring is often taken to be directly proportional to the disk viscosity parameter \citep{2007Icar..192..588Y}: $\mathcal{D}_{\rm{turb}}=\nu/\mathrm{Sc}=\alpha\,\Omegak\,h^2/\rm{Sc}$, where $\rm{Sc}$ is the turbulent Schmidt number\footnote{The turbulent Schmidt number is a measure of turbulent viscosity relative to the associated turbulent mixing.}. 

In the limit where the cumulative dust mass is negligible, competition among these two processes determines the thickness of the dust layer, and the expression for the solid sub-disk scale-height has the form \citep{1995Icar..114..237D}:
\begin{align}
h_{\bullet}=\frac{h}{\sqrt{1+\mathrm{Sc}\,\tau^{(\rm{eq})}/\alpha}}.
%\bigg(1+\frac{\alpha}{Sc\,\tau^{(\rm{eq})}}\bigg)^{-1/2}\sim h,
\label{eqn:hs}
\end{align}
Note that in the regime where $\alpha \ll \mathrm{Sc}\,\tau^{(\rm{eq})}$, we have $h_{\bullet}\ll h$, and equation (\ref{eqn:hs}) simplifies to the oft-quoted result $h_{\bullet}\approx h\,\sqrt{\alpha/(\tau^{(\rm{eq})}\,\rm{Sc}})$. Our model circumplanetary disk, however, lies at the opposite extreme of parameter space. The dust layer's aspect ratio ($h_{\bullet}/r$) for our fiducial parameters, computed in the massless particle limit quoted above, is shown as a black curve on the top panel of Figure (\ref{F:hr}). Because $\alpha \gg \tau^{(\rm{eq})}$, the vertical extent of the dust layer is comparable to that of the gas disk everywhere in the nebula, meaning that purely hydrodynamic settling of dust is exceedingly inefficient within the context of our model. Moreover, because $\tau^{(\rm{eq})}\propto1/\Sigma\propto\alpha/\dot{M}$ is linearly proportional to the disk viscosity, this determination is completely independent of the assumed value of $\alpha$.

A similar analysis can be undertaken for radial diffusion of aerodynamically trapped dust particles \citep{Dullemond2018}. As a representative example, consider a particle that satisfies the equilibrium equation (\ref{eqn:equilibrium}) at the reference radius, $r_0$. For our nominal parameters with $(1+\mathcal{Z}_{\rm{mid}})\sim 1$, this particle has a radius $s_{\bullet}\approx0.3\,$mm and lies in the Epstein regime. Retaining this value of $s_{\bullet}$, the radial evolution equation (\ref{eqn:radial2}) can be linearized around $r=r_0$. Assuming that $(4\,\pi\,\Sigma_0\,r_0^2\,\eta\,\Omegak/\Mdot)^2\gg1$ (which is very well satisfied for our model), the linearized equation for the particle's radial velocity in the vicinity of equilibrium takes on a rather rudimentary form:
\begin{align}
v_{r\,\bullet}^{(\rm{eq})}\approx-\frac{\Mdot\,\big(r-r_0\big)}{4\,\pi\,\Sigma_0\,r_0^2}.
\label{eqn:vr0}
\end{align}

The fact that $v_{r\,\bullet}^{(\rm{eq})}\propto-(r-r_0)$ demonstrates that the equilibrium is stable to perturbations, since aerodynamic drag exerts a restoring force on the particle. Adopting the disk viscosity for the turbulent diffusion coefficient as before\footnote{For this problem, it is appropriate to drop the reduction factor $\mathrm{Sc}$, since $\alpha$ is a viscosity parameter associated with radial angular momentum transport.}, and defining the variable $\xi=r-r_0; \dot{\xi}=v_{r\,\bullet}^{(\rm{eq})}$, we have the following stochastic equation of motion for the particle:
\begin{align}
d\xi= \alpha\,\Omega\,h^2\,d\mathscr{W} -\frac{\Mdot\,\xi}{4\,\pi\,\Sigma_0\,r_0^2}\,dt,
\label{eqn:dxi}
\end{align}
where $\mathscr{W}$ represents a drift-free Weiner process \citep{Oksendal}. An elementary result of stochastic calculus is that the solution to equation (\ref{eqn:dxi}) is a bounded random walk, with a characteristic dispersion
\begin{align}
\xi\sim h\,\sqrt{\frac{4\,\pi\,\alpha\,\Sigma_0\,r_0^2\,\Omegak}{\Mdot}}.
\label{eqn:dxians}
\end{align}

The quantity inside the square root exceeds unity by a large margin, implying that as long as the mass of the solid component of the system is negligibly small, dust within the circumplanetary disk is not only well-mixed vertically, but may also experience relatively long-range radial diffusion. In particular, for nominal disk parameters, $\xi\sim0.17\Rh$, implying that despite the functional form of the equilibrium (\ref{eqn:equilibrium}), diffusion prevents size-sorting of particles within the disk. 

Importantly, if we repeat this exercise in the Stokes regime (where $\tfric$ is given by equation \ref{eqn:Stokestfric}) for a different nominal radius $r_0'$, we find that $v_{r\,\bullet}^{(\rm{eq})}\approx7\,\Mdot\,\big(r-r_0'\big)/(8\,\pi\,\Sigma_0\,r_0^2).$ Because in this case $v_{r\,\bullet}^{(\rm{eq})} \propto +\,(r-r_0')$, a particle perturbed away from equilibrium will not experience a restoring radial acceleration, and will instead flow away from $r_0'$. This means that the derived equilibrium is only stable in the Epstein regime. Although a useful starting point, this discussion of stability along with equation (\ref{eqn:dxi}) should not be mistaken for a quantitative model of the radial distribution of particles, because the self-limiting cycle of grain growth and sublimation is likely to play an important dynamical role within the envisioned circumplanetary nebula.

This can be understood as follows: as particles coagulate to larger sizes, they experience stronger headwind (thus violating the equilibrium condition \ref{eqn:equilibrium}) and spiral-in towards the planet. Upon crossing the ice-line of the circumplanetary disk, however, sublimation ensues, causing the particle size to diminish, until it is small enough for the dust to be expelled back out by the decretion flow. Beyond the ice-line, grain growth ensues once again and the cycle repeats. Therefore, we expect that the particle distribution will be more strongly concentrated near the equilibrium radius than suggested by equation (\ref{eqn:dxi}) if we were to account for dust sublimation and growth. Moreover, because the equilibrium particle radius increases with $\mathcal{Z}_{\rm{mid}}$, for the successful operation of our scenario, we must envision that the coagulation/sublimation cycle operates considerably faster than the growth of the mid-plane metallicity, thus maintaining $s\sim s_{\bullet}^{(\rm{eq})}$ on average even as $\mathcal{Z}$ slowly increases.

%Let us begin be recalling Given that here we are concerned with turbulent transport of dust in the radial direction, it is appropriate to omit the Schmidt number.

%ALSO: REMARK ON THE GROWTH OF DUST WHICH LEADS TO INWARD MIGRATION, WHICH IN TURN LEADS TO ICY DUST MELTING AND GETTING EXPELLED BACK OUT. THE OPPOSITE EFFECT LIKELY OPERATES IN THE REVERSE DIRECTION AS WELL: PARTICLES THAT MOVE OUTWARDS AND ATTAIN COAGULATE TO LARGER SIZES, EXPERIENCE STRONGER INWARD MIGRATION DUE TO THEIR ENHANCED INTERACTIONS WITH THE AERODYNAMIC DRAG.

\section{Satellitesimal Formation} \label{sec:tesimal}

Our results from the previous section demonstrate that solid grains of a particular size-range can be captured within the circumplanetary nebula, by attaining a balance between aerodynamic drag and radial updraft associated with the decretion flow. However, these particles have long settling times, and do not readily sediment into a vertically confined sub-disk. Instead, as long as the overall metallicity, $\mathcal{Z}=\Sigmad/\Sigma$, of the circumplanetary disk remains very low, we can envision that the gaseous and solid components of the system continue to be well-mixed. Nevertheless, it is clear that this simple picture cannot persist indefinitely.

As the host planet continues to evolve within the protoplanetary disk, incoming meridional circulation and ablation of injected planetesimals act to slowly enrich the circumplanetary disk in dust. This occurs at a rate:
\begin{align}
\Mdot_{\bullet} = f_{\bullet}\,\Mdot+\big( \Mdot_{\bullet} \big)_{\rm{ablation}},
\label{eqn:Mdotsolid}
\end{align}
where $f_{\bullet}$ is the mass fraction of incoming particles with Stokes numbers corresponding to stationary values $\tau^{(\rm{eq})}$ encapsulated by the disk (and those that can grow to the appropriate equilibrium size before getting expelled from the disk). Although the precise value of $\Mdot_{\bullet}$ depends on the adopted system parameters (and is somewhat poorly constrained), it is worthwhile to notice that a sufficient amount of solid mass to build the Jovian and Kronian moons (i.e., $M_\bullet/\M\sim2\times10^{-4}$) can be accumulated in a million years, even if planetesimal ablation is completely neglected and the mass-fraction of incoming $\sim0.1-1\,$mm particles is assumed to be a mere $f_{\bullet}\sim0.002$ -- about an order of magnitude smaller than the usual dust-to-gas ratio of protoplanetary disks \citep{Armitage2010}. 

In light of the fact that in reality, $\Mdot$ can be considerably larger than our fiducial value of $0.1\M/$Myr at earlier epochs \citep{Lambrechts2019}, and that particle injection from planetesimal ablation can further increase the rate of dust buildup within the system \citep{2010SSRv..153..431M,RonnetJohansen2020}, the aforementioned estimate almost certainly represents a gross lower bound on the actual amount of solid material that can be effectively sequestered in the disk. Moreover, as the parent (circumstellar) nebula gradually fades, the gaseous component of the circumplanetary disk must also diminish in time due to a decreasing $\Mdot$, thus gradually enhancing $\mathcal{Z}$. As a result, it seems imperative to consider the possibility that a circumplanetary disk can readily approach a strongly super-solar metallicity (perhaps even of order unity) during its lifetime. Accordingly, let us examine the vertical distribution of dust in a circumplanetary disk with $\mathcal{Z}\lesssim1$.

Independent of the degree of coupling that solid particles experience with the gas, an inescapable limitation on the vertical dust stirring that turbulence can facilitate lies in the energy budget of this process. That is, in order to lift the dust above the mid-plane, turbulent eddies must do gravitational work, and the available kinetic energy to do this work is necessarily restricted. Under the assumptions that underly equation (\ref{eqn:alpha}), characteristic velocity of turbulent eddies that manifest at the largest scales ($\ell\sim\sqrt{\alpha}\,h$) is on the order of $v_{\rm{turb}}\sim\,\sqrt{\alpha}\,\cs$. Thus, the column-integrated turbulence kinetic energy density is:
\begin{align}
\mathcal{E}_{\rm{turb}}\sim\int_{-\infty}^{\infty}\frac{1}{2}\,\rho\,v_{\rm{turb}}^2\,dz = \frac{\alpha}{2}\,\Sigma\,\cs^2.
\label{eqn:Ekinetic}
\end{align}

When dust is lifted above the disk mid-plane, this kinetic energy gets converted into gravitational potential energy. In a geometrically thin disk, the downward gravitational acceleration experienced by dust can be approximated as $g\approx\Omega^2\,z$ \citep{Armitage2010}. Assuming that the solid component of the disk follows a Gaussian profile like the gas, the column-integrated gravitational potential energy of the dust layer has the form:
\begin{align}
\mathcal{E}_{\rm{grav}}\sim\int_{-\infty}^{\infty}\rho_{\bullet}\,g\,z\,dz = \Sigma_{\bullet}\,\Omegak^2\,h_{\bullet}^2.
\label{eqn:Grav}
\end{align}
Setting $\epsilon\,\mathcal{E}_{\rm{turb}}=\mathcal{E}_{\rm{grav}}$, we arrive at the energy-limited expression for the dust-layer's aspect ratio:
\begin{align}
\boxed{\frac{h_{\bullet}}{r}\sim\sqrt{\frac{\alpha\,\epsilon}{2\,\mathcal{Z}}}\,\bigg(\frac{h}{r}\bigg),}
\label{eqn:hrenergy}
\end{align}
where $\epsilon \leqslant 1$ is a numerical factor that accounts for the fact that only a fraction of the vertically integrated turbulence kinetic energy goes into elevating the dust above the mid-plane, as well as for the suppression of turbulence by enhanced dust concentration \citep{2019MNRAS.485.5221L}. For definitiveness, in this work we adopt $\epsilon=0.1$, but remark that this guess is highly uncertain. 

The physical meaning of equation (\ref{eqn:hrenergy}) can be understood in a straight-forward manner: even if dust within the circumplanetary disk is sufficiently well coupled to the gas for it to potentially remain well-mixed throughout the vertical extent of the disk, enhancing the scale-height of the solid sub-disk by turbulent stirring comes at a steep energetic cost. As a result, in relatively quiescent, dust-rich systems (where $\alpha\ll\mathcal{Z}$), the solid layer will necessarily be thin compared to the gas disk\footnote{This effect highlights yet another important distinction between the physics of satellite formation and planet formation. In typical protoplanetary disks, $\mathcal{Z}\sim0.01$ and dust sedimentation towards the mid-plane occurs simply due to the fact that for a broad range of particle sizes, $\alpha \ll \tau^{(\rm{eq})}$ (see equation \ref{eqn:hs}).}. Proceeding under the assumption of a Gaussian profile as before, we may readily write down the functional form of the mid-plane metallicity by taking the ratio of mid-plane densities:
\begin{align}
\mathcal{Z}_{\rm{mid}} = \frac{\Sigma_{\bullet}}{\Sigma}\frac{h}{h_{\bullet}}\sim\mathcal{Z}\sqrt{\frac{2\,\mathcal{Z}}{\alpha\,\epsilon}}.
\label{eqn:Zmid}
\end{align}
Noting that $\alpha\,\epsilon \sim 10^{-5}$, we remark that in order for $\mathcal{Z}_{\rm{mid}}$ to exceed unity, the overall disk metallicity must only exceed the solar value by a factor of a few. The ensuing vertical confinement of dust has profound consequences for formation of satellitesimals. 

By now, it is well-established that a broad range of gravito-hydrodynamic instabilities can develop within two-fluid mixtures of gas and dust \citep{YoudinGoodman2005,2007Natur.448.1022J,2018ApJ...856L..15S,Seligman2019}. These remarkable phenomena, however, only emerge at sufficiently high (local) concentration of high-metallicity material. A broadly discussed example of this group of instabilities is known as the streaming instability \citep{YoudinGoodman2005,JohansenYoudin2007}, which can facilitate rapid growth of dust clouds within protoplanetary disks through a back reaction of accumulated solid particles on the background quasi-Keplerian flow. A distinct variant of a two-fluid instability is known as the $\mathcal{Z}>1$ resonant drag instability \citep{2018ApJ...856L..15S} and can also promote the coagulation of solid material within the disk, albeit at smaller scales. Importantly, within the context of the protosolar nebula, it is now widely speculated that enhancement of the local solid-to-gas ratio associated with the aforementioned effects can culminate in gravitational collapse of particle clouds, resulting in the formation of bonafide planetesimals.

The emergence of gravito-hydrodynamic resonant drag instabilities within dust-loaded circumplanetary decretion disks is an intriguing possibility that deserves careful investigation with the aid of high-resolution numerical simulations. At the same time, this exercise falls beyond the immediate scope of our (largely analytic) study. Accordingly, to circumvent this riveting complication, here we focus our attention on the qualitatively simplest pathway for conversion of dust into satellitesimals: sedimentation, followed by direct gravitational collapse i.e., the Goldreich-Ward mechanism (\citealt{GoldreichWard1973}; see also \citealt{YoudinShu2002}). 

\begin{figure}[tbp]
\centering
\includegraphics[width=\columnwidth]{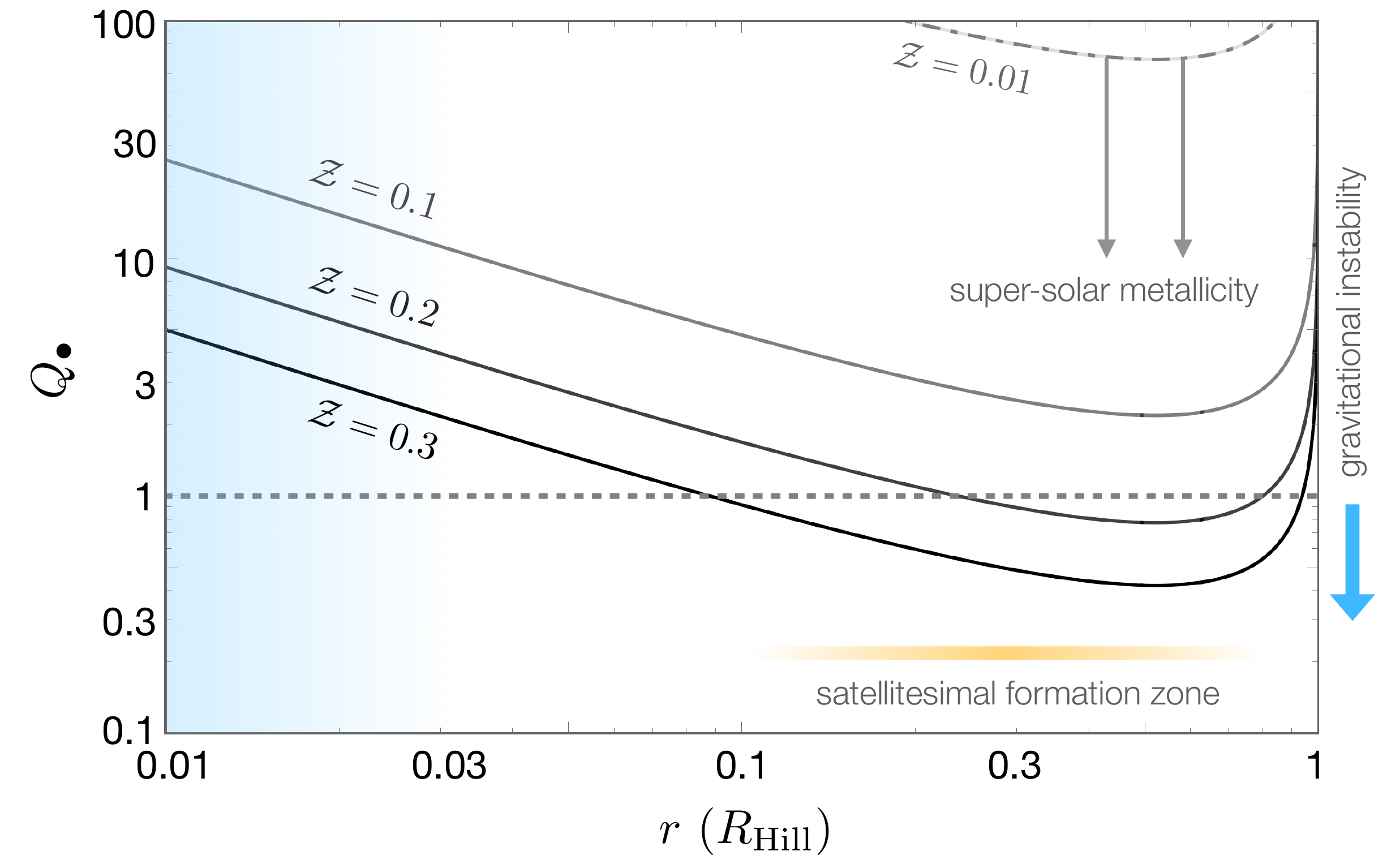}
\caption{Gravitational stability of the solid sub-disk. Toomre's Q parameter (equation \ref{eqn:Q}) is shown as a function of orbital radius, for a sequence of circumplaentary nebula metallicities. For solar composition gas with $\mathcal{Z}=0.01$, $Q_{\bullet}\gg 1$, and the solid sub-disk is gravitationally stable. However, for dust-to-gas ratio in excess of $\mathcal{Z}\gtrsim0.2$, the solid sub-disk becomes gravitationally unstable, fragmenting into satellitesimals. Notably, for $\mathcal{Z}=0.3$ -- which we take as a reasonable estimate for the onset of large-scale satellitesimal formation -- gravitational collapse can ensue outwards of $r\gtrsim0.1\,\Rh$.}
\label{F:Q} 
\end{figure}

Linear stability analysis of differentially rotating self-gravitating disks carried out over half a century ago \citep{Safronov1960,Toomre1964}, has shown that in order for gravitational collapse to ensue in presence of Keplerian shear, the system must satisfy the following rudimentary criterion:
\begin{align}
Q_{\bullet}=\frac{h_{\bullet}\,\Omegak^2}{\pi\,\G\,\Sigma_{\bullet}}=\frac{h\,\Omegak^2}{\pi\,\G\,\Sigma}\sqrt{\frac{\alpha\,\epsilon}{2\,\mathcal{Z}^3}}\lesssim1.
\label{eqn:Q}
\end{align}
For our adopted benchmark parameters of $\Mdot=0.1\,\M/$Myr and $\alpha=10^{-4}$, this expression dictates that direct conversion of dust into satellitesimals can be triggered within the circumplanetary disk for $\mathcal{Z}\gtrsim0.2$ (Figure \ref{F:Q}). Equations (\ref{eqn:powerlaw}) indicate that $Q_\bullet\appropto r^{-3/4}$, implying that like in the case of circumstellar nebulae (see e.g., \citealt{Boss1997}), gravitational collapse is more easily activated in the outer regions of our model circumplanetary disk. Notably, this preference for longer orbital periods for generation of satellitesimals is generic, and would apply even if pre-collapse agglomeration of solids is assisted by some two-fluid instability (\citealt{Yang2017} and the references therein).

It is well known that our envisioned process for satellitesimal formation (where dust consolidates directly into satellitesimals under its own gravity) can be suppressed under certain conditions in real astrophysical disks. To this end, \citet{Goldreich2004} point out that in order for gravitational fragmentation to ensue, the particle disk must be optically thick. Quantitatively, this criterion translates to $\Sigma_{\bullet}/(\rho_{\bullet}\,s_{\bullet})\gtrsim1$. This limit does not pose an issue for the problem at hand, because the smallness of equilibrium particle size within our disk (as dictated by equation \ref{eqn:equilibrium}) ensures that this inequality trivially satisfied. 

A more acute suppression mechanism for the Goldreich-Ward instability is the turbulent self-regulation of dust settling. That is, as the dust layer is envisioned to grow thinner, its mid-plane azimuthal velocity inevitably approaches the purely Keplerian value. The shear associated with the development of a vertical gradient in $v_{\phi}$ gives rise to the Kelvin-Helmholtz instability, which turbulently stirs the dust layer, counteracting sedimentation \citep{1980Icar...44..172W,Cuzzi1993}. While this process can indeed subdue planetesimal formation in the circumstellar nebula where $\mathcal{Z}\sim0.01$, for the circumplanetary system at hand, this problem is circumvented by virtue of the disk having a sufficiently high metallicity. In other words, even if the solid sub-disk is perturbed by turbulence, the dust cannot be lifted appreciably due to energetic limitations (equation \ref{eqn:hrenergy}). This reasoning is supported by the results of \citet{1998Icar..133..298S}, who demonstrated that for $\mathcal{Z}\gtrsim0.1$, dust stirring becomes inefficient, allowing gravitational collapse to proceed even in presence of parasitic Kelvin-Helmholtz instabilities. Consequently, we conclude that while the development of Kelvin-Helmholtz instabilities can suppress the Goldreich-Ward mechanism in the protosolar nebula, direct gravitational collapse of solid grains into planetesimals is possible in circumplanetary disks because of dust-loading within the system.

The dispersion relation associated with a self-gravitating Keplerian particle fluid has the well-known form \citep{Armitage2010}:
\begin{align}
\omega^2=c_{\bullet}^2\,k^2-2\,\pi\,\G\,\Sigma_{\bullet}\,|k|+\Omega^2,
\label{eqn:dispersion}
\end{align}
where $c_{\bullet}=(h_{\bullet}/r)\,\vk$ is the velocity dispersion of the dust. The critical wavenumber corresponding to the most rapidly growing unstable mode of this relation is
\begin{align}
k_{\rm{crit}}=\frac{\pi\,\G\,\Sigma_{\bullet}}{c_{\bullet}^2}\sim\frac{1}{h_{\bullet}},
\label{eqn:kcrit}
\end{align}
where the RHS follows from setting $Q_{\bullet}\sim1$ in equation (\ref{eqn:Q}). Accordingly, the characteristic mass scale of planetesimals generated through gravitational collapse is:
\begin{align}
\boxed{m\sim\pi\,\bigg(\frac{2\,\pi}{k_{\rm{crit}}} \bigg)^2\Sigma_{\bullet}=2\,\pi^3\,\alpha\,\epsilon\,h^2\,\Sigma.}
\label{eqn:mcrit}
\end{align}
Importantly, to arrive at the RHS of this estimate, we have used expression (\ref{eqn:hrenergy}) for $h_{\bullet}$ to cancel out the dependence on $\mathcal{Z}$.

The fact that this expression is independent of $\mathcal{Z}$ is qualitatively important. Indeed, while the metallicity dictates whether or not gravitational collapse can be triggered via equation (\ref{eqn:Q}), the physical properties of satellitesimals generated through fragmentation of the solid sub-disk are largely determined by the global properties of the circumplanetary nebula. For our fiducial parameters, we obtain bodies with $m\sim10^{19}\,$kg at $r\sim0.1-0.3\,\Rh$ -- comparable to the mass of Mimas, and about two orders of magnitude smaller than the mass of Iapetus. For a mean density of 1\,g/cc, this mass scale translates to bodies with radii on the order of $100\,$km.

\section{Satellite Growth and Migration} \label{sec:growth}

While the characteristic mass-scale of satellitesimals generated through gravitational instability is appreciable, it is still negligible compared to the cumulative mass of the Galilean moons or Titan, meaning that additional growth must take place to explain the satellite systems of the giant planets. Growth of solid bodies within the circumplanetary disk can proceed via two potential pathways: pairwise satellitesimal collisions or pebble accretion. Recently, the pebble accretion paradigm has been shown to be remarkably successful in resolving long-standing issues of planet-formation \citep{2010A&A...520A..43O,LambrechtsJohansen2012,2012A&A...546A..18M,RonnetJohansen2020}, and has consequently gained considerable traction within the broader community. Inspired by this mechanism's growing fashionableness, let us begin this section by considering growth of satellitesimals by pebble capture within the circumplanetary disk.

\subsection{Pebble Accretion} \label{sec:pebble}

Depending on the mass of the growing satellitesimal and the degree of dust-gas coupling, its propensity towards aerodynamically assisted capture of small particles can proceed in one of two modes of accretion: the Bondi regime or (the considerably more efficient) Hill regime. In the former case, the key physical length and time scales that characterize the pebble accretion process are the Bondi radius, and the corresponding crossing time:
\begin{align}
&\Rb=\frac{\G\,\mathcal{M}}{\Delta v^2} &t_{\rm{B}}=\frac{\Rb}{\Delta v},
\label{eqn:Rbondi}
\end{align}
where $\mathcal{M}$ is the satellite embryo's mass, and $\Delta v=\eta\,\vk/(1+\mathcal{Z}_{\rm{mid}})$ is the particle approach speed. Qualitatively, $\Rb$ represents a critical impact parameter below which the gravitational potential of the growing body can facilitate large-angle deflection of dust, and $t_{\rm{B}}$ is the characteristic timespan associated with the encounter. %Conveniently, for $\tau\ll 1$, the particle relative drift speed is $\Delta v=\eta\,\vk/(1+\mathcal{Z}_{\rm{mid}})$ to an excellent approximation \citep{LambrechtsJohansen2012}.

Contrary to the case of the protosolar nebula, where a broad size distribution of dust particles translates to an extended range of Stokes numbers, the aerodynamic equilibrium delineated in section \ref{sec:dust} ensures that the circumplanetary disk is loaded with solid particles that are characterized by a similar frictional timescale, $\tfric^{(\rm{eq})}$. This allows us to define an almost unique capture radius for circumplanetary dust in the Bondi regime (\citealt{2010A&A...520A..43O}):
\begin{align}
&\Rc\approx\Rb\sqrt{\frac{\tfric}{t_{\rm{B}}}}=\frac{\G\,\mathcal{M}\,\big(1+\mathcal{Z}_{\rm{mid}} \big)}{\eta\,\vk^2}\sqrt{\frac{r\,v_r\,\vk}{2\,\G\,\mathcal{M}}},
\label{eqn:Rc}
\end{align}
where we have used equation (\ref{eqn:equilibrium}) to relate $\tfric$ to $v_r$.

In the Hill regime, the effective capture radius is \citep{2016A&A...591A..72I}:
\begin{align}
\Rch&\approx r\,\bigg(\frac{10\,\tau\,\mathcal{M}}{3\,\M} \bigg)^{\frac{1}{3}} =r\,\bigg( \frac{5\,v_r\,\mathcal{M}\,(1+\mathcal{Z}_{\rm{mid}}) }{3\,\eta\,\vk\,\M}\bigg)^{\frac{1}{3}} .
\label{eqn:Rch}
\end{align}
The crossover between the two modes of accretion occurs when the effective Bondi and Hill accretion radii are equivalent i.e., $\Rc\sim\Rch$. After some rearrangement (see appendix \ref{appndB}), this yields a transitionary embryo mass of order:
\begin{align}
\mathcal{M}_{\rm{t}}\sim\frac{400\,\pi}{9}\,\bigg(\frac{\eta}{1+\mathcal{Z}_{\rm{mid}} } \bigg)^4\, \bigg( \frac{\Sigma\,r^2\,\Omega}{\Mdot} \bigg)\,\M.
\label{eqn:mtrans}
\end{align}
If we assume that conversion of dust into solid bodies is less than 100\% efficient, such that following large-scale satellitesimal formation the pebble surface density is still on the order of $\Sigma_{\bullet}/\Sigma\sim0.1$ (which corresponds to $\mathcal{Z}_{\rm{mid}}\approx14$), the above expression suggests that the Hill regime of accretion is appropriate for satellites more massive than $\mathcal{M}\gtrsim \mathcal{M}_{\rm{t}}\sim\rm{few}\,\times10^{24}\,$kg. This mass-scale exceeds the mass of Ganymede by more than an order of magnitude, implying that any accretion of pebbles in our model circumplanetary disk is sure to proceed in the Bondi regime.

%where the pebble approach speed is $\Delta v=\eta\,\vk/(1+\mathcal{Z}_{\rm{mid}})$. For $(1+\mathcal{Z}_{\rm{mid}})\approx75$ -- appropriate for $\mathcal{Z}\approx0.3$ (which we take as an estimate for the onset of large-scale satellitesimal formation) -- the transition mass evaluates to $m_{\rm{t}}\sim2\,\times\,10^{17}\,$kg. This mass scale is more than an order of magnitude smaller than the typical satellitesimal mass given by equation (\ref{eqn:mcrit}), meaning that upon formation via gravitational collapse, satellitesimals can begin accreting pebbles directly in the Hill regime.

For $\mathcal{M}\lesssim10^{22}\,$kg, $\Rc\lesssim h_{\bullet}$. Recalling that the characteristic mass-scale of satellitesimals that form via gravitational collapse is $m\sim10^{19}\,$kg, this means that the dust layer is much more vertically extensive than the pebble capture radius, implying that accretion unfolds in 3D. The corresponding accretion rate has the form (see appendix \ref{appndB} for additional details)
%Within the context of Hill-limited accretion, the mass growth rate is given by \citep{2010A&A...520A..43O,LambrechtsJohansen2012}:
\begin{align}
&\bigg(\frac{d\,\mathcal{M}}{d\,t}\bigg)_{\rm{B \, 3D}} = \sqrt{\frac{\pi}{2}}\,\frac{\Sigma_{\bullet}}{h_{\bullet}}\,\big( \Rc \big)^2\,\Delta v \nonumber \\
&=\frac{1}{2}\frac{\mathcal{M}}{\M}\frac{1+\mathcal{Z}_{\rm{mid}}}{\eta}\bigg( \frac{h}{r}\bigg)^{-1}\sqrt{\frac{\pi\,\mathcal{Z}^3}{\alpha\,\epsilon}}\,\Sigma\,r\,v_r,
\label{eqn:BONDI3D}
\end{align}
where we have adopted $\Sigma_{\bullet}/(\sqrt{2\,\pi}\,h_{\bullet})$ as an estimate for the volumetric density of the pebble disk. Setting $\mathcal{Z}\sim0.1$, this formula yields a mass-doubling time of $(1/\mathcal{M}\,\times\,d\,\mathcal{M}/d\,t)^{-1} \sim2{,}000$ years.

While the above estimate of relatively rapid accretion may appear promising, it is unlikely that it translates to significant long-term growth. This is because within the context of our model, pebble accretion is necessarily limited by the satellitesimal's access to the overall supply of dust. That is, unlike the oft-considered case of a circumstellar accretion disk, where inward drift of pebbles acts to refill the orbital neighborhood of growing planetesimals, in the circumplanetary disk, the aerodynamic equilibrium discussed in section \ref{sec:dust} implies that no steady-state drift exists. Instead, here radial dispersion of pebbles is driven almost entirely by turbulence viscosity, yielding diffusion-limited growth of satellitesimals that is reminiscent of planetesimal growth within dust-loaded pressure bumps (recently considered by \citealt{Morbidelli20202}). Thus, crudely speaking, the reservoir of solid dust that is available to any given satellitesimal is restricted to the material that is entrained between the satellitesimal itself and its nearest neighbors.

If we envision that large-scale gravitational collapse of the solid sub-disk yields a population of debris that is comparable in total mass to that of the remaining dust disk, and that the generated satellitesimals commence their growth at approximately the same time, the above reasoning implies that the pebble accretion process can only boost the individual masses of satellitesimals by a factor of $\sim2$ before the global supply of dust is exhausted. We therefore conclude that within the context of our model, pebble accretion can only yield a short-lived burst of satellitesimal growth, and is unlikely to be the dominant mechanism for converting satellitesimals into full-fledged satellites. In light of the short-lived nature of this process, coupled with considerable uncertainties on the efficiency its operation, we will neglect it for the remainder of the paper.

\subsection{Oligarchic Growth} \label{sec:olig}

If pebble accretion is ineffective in boosting the masses of satellitesimals by an appreciable amount, long-term conglomeration of the large giant planet satellites must occur through pairwise collisions\footnote{Notably, this mode of accretion is qualitatively much closer to the standard picture of terrestrial planet formation than it is to the emergent picture of the formation of Super-Earths.}. In this case, the rate of accretion experienced by a satellite embryo is dictated by an $n$-$\sigma$-$v$ type relation, and has the form \citep{Lissauer1993}:
\begin{align}
\frac{d\,\mathcal{M}}{d\,t} = 4\,\pi\,\bar{\rho}\,\R^2\,\frac{d\,\R}{d\,t}= \kappa\,\Lambda\,\mathcal{Z}\,\Sigma\,\,\pi\,\R^2(1+\Theta)\,\Omegak,
\label{eqn:dMdt}
\end{align}
where $\Lambda\leqslant1$ is the efficiency of conversion of dust into satellitesimals through gravitational instability, $\kappa$ is a constant of order unity\footnote{For an isotropic velocity dispersion, $\kappa=\sqrt{3}/2$ \citep{Lissauer1993}.}, and $\Theta=(v_{\rm{esc}}/\langle v \rangle)^2$ is the Safronov number. Qualitatively, the parameter $\Lambda$ regulates the total mass of the satellitesimal swarm in the region where $Q_{\bullet}\lesssim1$. Given that neither the satellitesimal generation process nor the satellite accretion process are expected to be perfectly efficient, a value of $\Lambda\sim1/3-1/2$ appears reasonable within the framework of our model. We note, however, that any value of $\Lambda$ in excess of $\sim0.1$ yields a debris disk between $0.1\,\Rh$ and $0.3\,\Rh$ that exceeds the total mass of the observed satellites.

The solution to equation (\ref{eqn:dMdt}) for $\R$ as a function of $t$ is trivially obtained if the mean density and the parameters on the RHS of the differential equation are assumed to be time-invariant. For the purposes of the following discussion, it is instructive to recast this solution in terms of an accretion timescale, $\mathcal{T}_{\rm{accr}}$, corresponding to a change in the embryo's radius from $\R_0 \rightarrow \R$:
\begin{align}
%\R=\R_0+\frac{\Lambda\,\mathcal{Z}\,(1+\Theta)\,\Sigma\,\Omega}{4\bar{\rho}}\mathcal{T}_{\rm{accr}}
\mathcal{T}_{\rm{accr}} = \frac{4\,\bar{\rho}\,(\R-\R_0)}{\kappa\,\Lambda\,\mathcal{Z}\,(1+\Theta)\,\Sigma\,\Omegak}.
\label{eqn:Taccr}
\end{align}

\begin{figure}[tbp]
\centering
\includegraphics[width=\columnwidth]{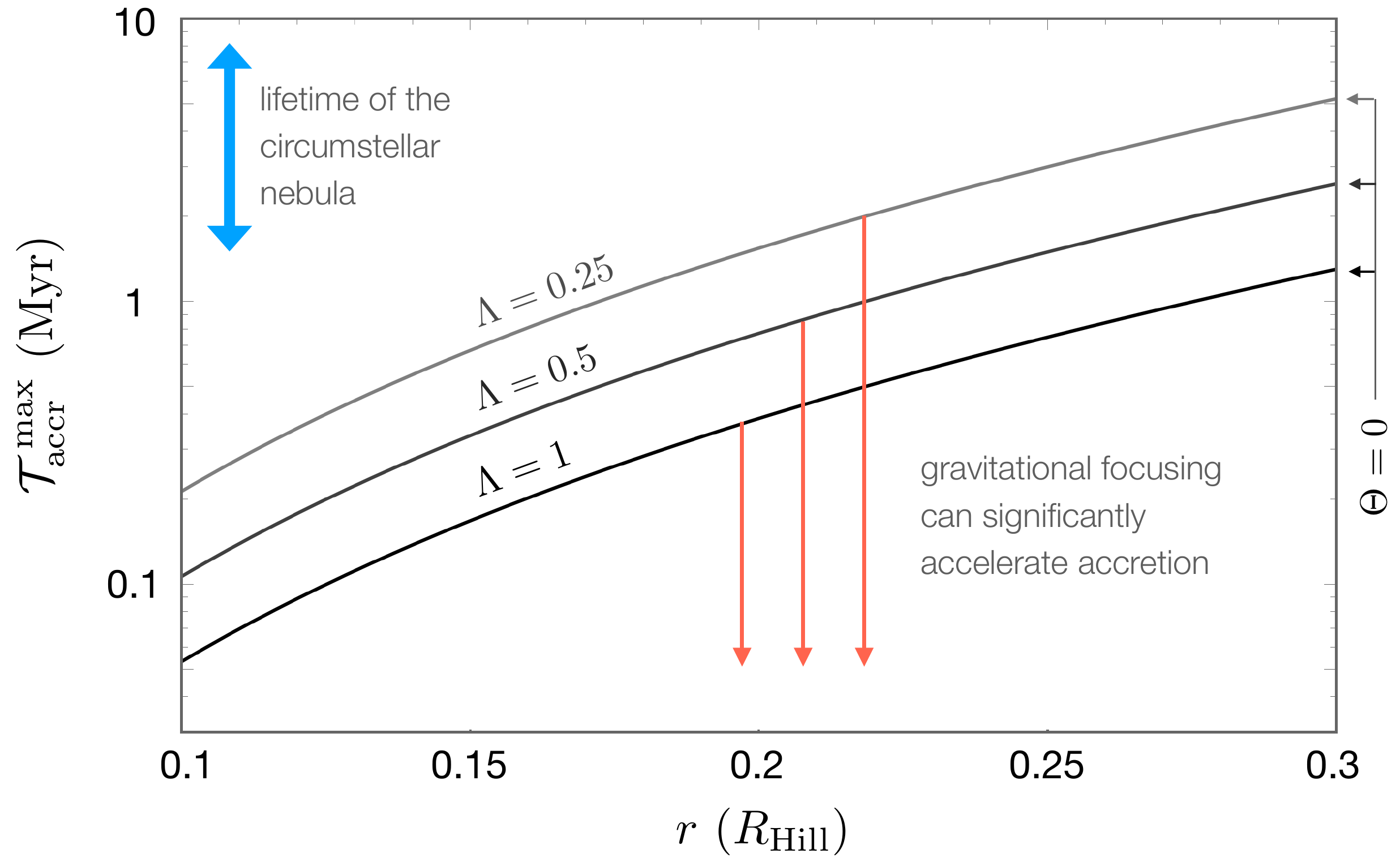}
\caption{Maximal accretion timescale of satellites (equation \ref{eqn:Taccr}). The depicted curves correspond to disk metallicity of $\mathcal{Z}=0.3$, and satellitesimal generation efficiency of 100\% ($\Lambda=1$), 50\% ($\Lambda=0.5$), and 25\% ($\Lambda=0.25$). The mean density and terminal radius are taken to be $\bar{\rho}=2\,$g/cc and $\mathcal{R}=2600\,$km, respectively. While we assume the Safronov number, $\Theta$, to be null for the purposes of this figure, it is important to keep in mind that gravitational focusing can significantly accelerate satellite formation, if the satellitesimal velocity dispersion is low. Note further that the accretion rate is semi-major axis-dependent, and proceeds more than an order of magnitude faster at $r\sim0.1\Rh$ than at $r\sim0.3\Rh$.}
\label{F:Taccr} 
\end{figure}

By neglecting gravitational focusing in the above expression (setting $\Theta\rightarrow0$), we can obtain an approximate upper limit on the formation timescale of large satellites, $\mathcal{T}_{\rm{accr}}^{\rm{max}}$, as a function of $r$ in our model disk. Retaining the same system parameters as those delineated in the proceeding sections, and setting $\R=2600\,$km; $\bar{\rho}=2\,$g/cc (approximate radius and mean density of Ganymede), we show $\mathcal{T}_{\rm{accr}}^{\rm{max}}$ in Figure (\ref{F:Taccr}) for $\Lambda=1,1/2$ and $1/4$. Crucially, this result demonstrates that the characteristic conglomeration timescale can reasonably exceed a million years, provided that the velocity dispersion of satellitesimals exceeds the escape velocity of the satellite embryo. 

In the opposite limit where the satellitesimal swarm is taken to be (initially) dynamically cold, embryo growth can proceed on a much shorter timescale, at first. However, the satellitesimal swarm cannot remain dynamically cold indefinitely, due to self-stirring and interactions with the accreting embryo -- a caveat best addressed with the aid of detailed simulations. In any case, the accretion process necessarily stops once the newly-formed satellite is ejected from the annulus of the circumplanetary disk occupied by the satellitesimal swarm. Within the context of our scenario, we envision this to occur as as consequence of satellite-disk interactions \citep{GoldreichTremaine1980,Ward1997}, which sap the satellite of its orbital angular momentum, leading to its progressively rapid orbital in-spiral. The characteristic timescale for the satellite's inward (type-I) migration is given by \citep{Tanaka2002}:
\begin{align}
%\R=\R_0+\frac{\Lambda\,\mathcal{Z}\,(1+\Theta)\,\Sigma\,\Omega}{4\bar{\rho}}\mathcal{T}_{\rm{accr}}
\mathcal{T}_{\rm{mig}} = \frac{\gamma}{\Omegak}\,\frac{\M}{\mathcal{M}}\,\frac{\M}{\Sigma\,r^2}\,\bigg(\frac{h}{r}\bigg)^2,
\label{eqn:Tmig}
\end{align}
where $\gamma$ is yet another dimensionless constant of order unity.

Importantly, expression (\ref{eqn:Tmig}) states that the migration timescale is inversely proportional to the satellite mass. This means that long-range orbital decay cannot ensue until the satellite is sufficiently large. Consequently, to obtain a crude limit on $\R$ (or $\mathcal{M}$) we follow \citet{CanupWard2002,CanupWard2006} and set $\mathcal{T}_{\rm{mig}}\sim\mathcal{T}_{\rm{accr}}$. After some rearrangement, we have
\begin{align}
\boxed{\R^3\,(\R-\R_0)\sim\frac{3\,\kappa\,\gamma}{16\,\pi}\,\bigg(\frac{h\,\M}{\bar{\rho}\,r^2}\bigg)^2\,\mathcal{Z}\,\Lambda\,(1+\Theta).}
\label{eqn:TmigTaccr}
\end{align}
While a closed form expression for $\R$ does exist, it is cumbersome, and does not elucidate any physics that is not already evident upon inspection of equation (\ref{eqn:TmigTaccr}). Accordingly, here we limit ourselves to simply noting that the solution for $\R$ is roughly given by the fourth root of the RHS of the above expression, and that this approximation improves for larger $\R$ (recall that $\R_0$ is set by the typical satellitesimal mass, given by equation \ref{eqn:mcrit}).

\begin{figure}[tbp]
\centering
\includegraphics[width=\columnwidth]{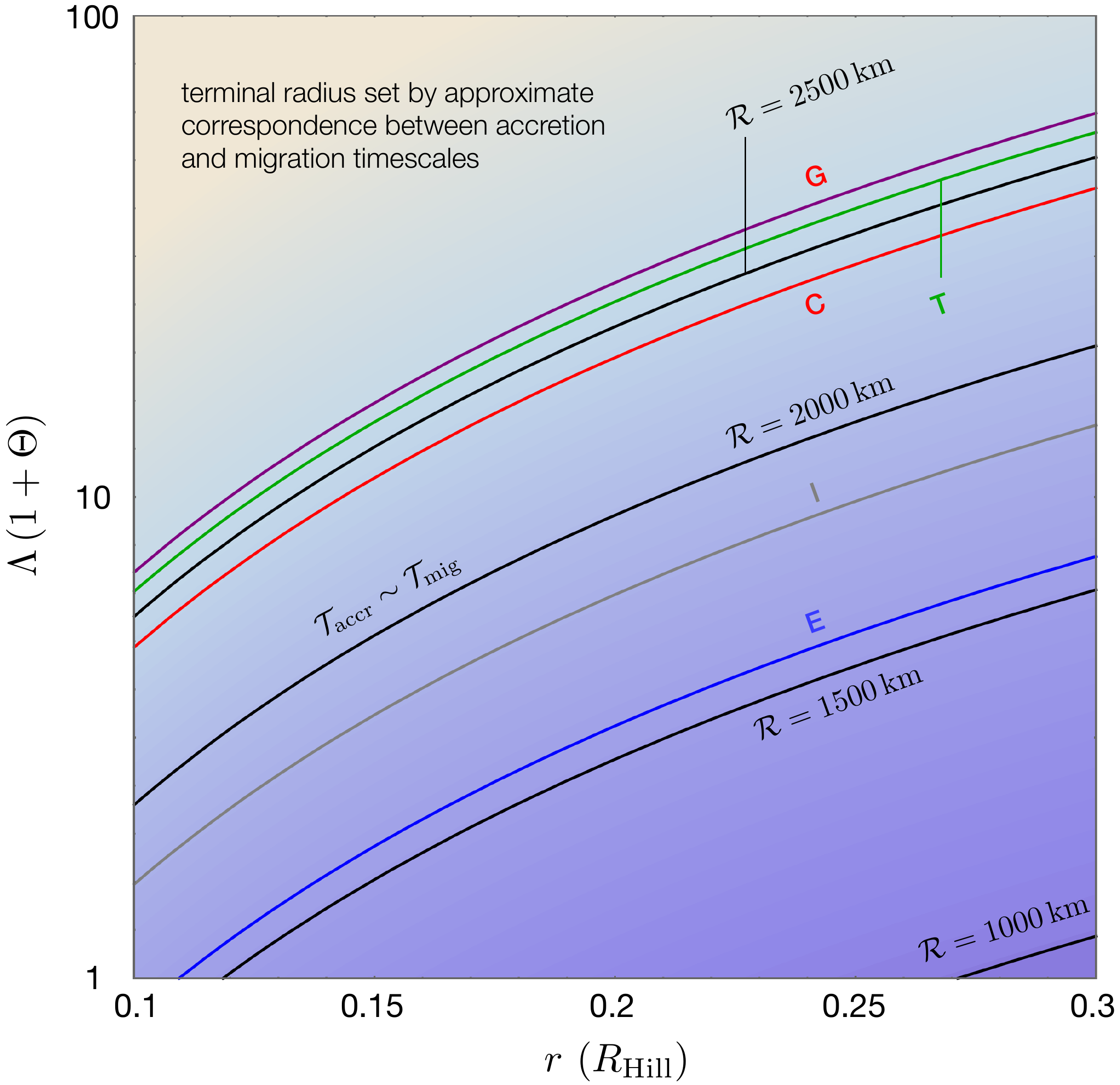}
\caption{Terminal radii of satellites. Within the framework of our model, satellite conglomeration continues until long-range orbital decay ensues, and removes the growing embryo from its $r\gtrsim0.1\Rh$ feeding zone. Accordingly, the terminal mass (and radius) of a forming satellite is approximately determined by equating the (type-I) migration timescale to the accretion timescale (equation \ref{eqn:TmigTaccr}). Contours of terminal satellite radii equal to 1000\,km, 1500\,km, 2000\,km and 2500\,km are shown as black curves on the figure. Contours corresponding to the true radii of the Galilean satellites and Titan are depicted with colored lines, and are labeled. A disk metallicity of $\mathcal{Z}=0.3$ and a mean satellite density of $\bar{\rho}=2\,$g/cc are assumed. The precise values of order-unity constants $\kappa$ and $\gamma$ are ignored.}
\label{F:rF} 
\end{figure}

The above discussion indicates that the terminal radius of a satellite is determined by two parameters: the dynamical temperature of the satellitesimal swarm $\Lambda\,(1+\Theta)$ and the planetocentric radius $r$. Contours of terminal $\R$ are shown in Figure (\ref{F:rF}), with real satellite radii marked with colored lines. Remarkably, this rudimentary analysis suggests that satellites with radii in the range $\R\sim1500-2500\,$km, can be naturally generated within the circumplanetary decretion disk, provided that satellitesimal disk that forms them originates with a low velocity dispersion. 

\section{Numerical Experiments} \label{sec:sims}

Without a doubt, the actual process of satellite formation is more complicated than the narrative foretold by the simple calculations presented above. Accordingly, it is imperative that we examine the validity of the emerging picture with more detailed numerical simulations. This is the primary purpose of this section. 

\subsection{Accretion Calculation} \label{sec:boulder}

\begin{figure*}[tbp]
\centering
\includegraphics[width=\textwidth]{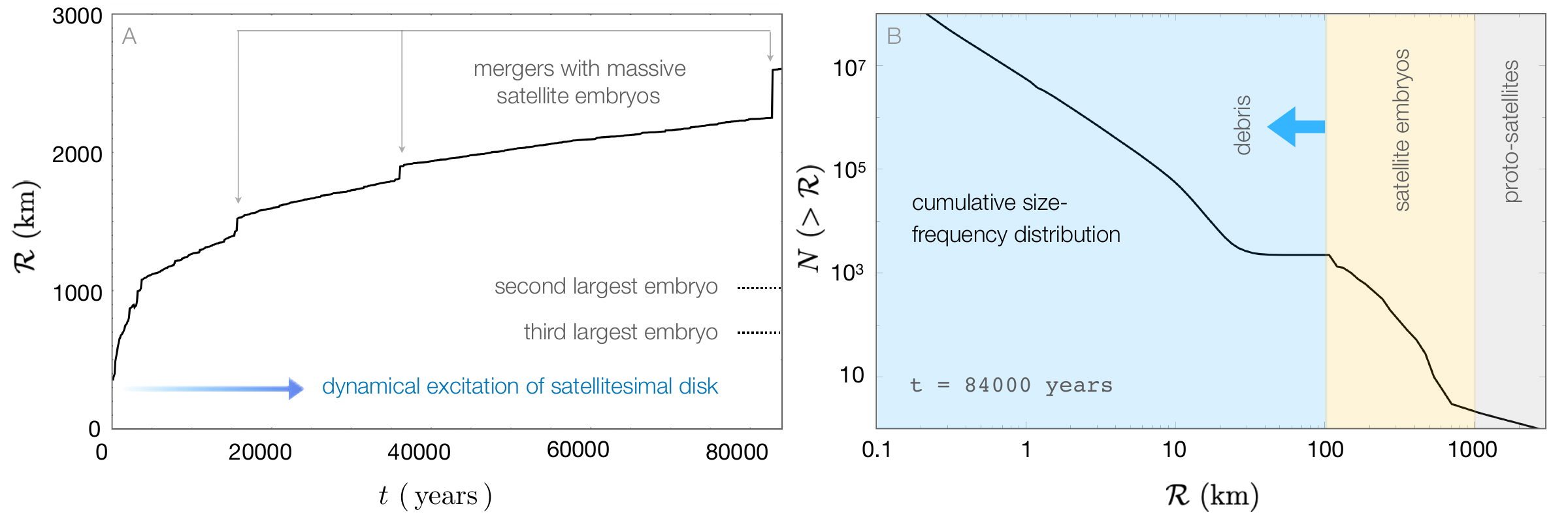}
\caption{Particle-in-a-box calculation of satellite accretion within the circumplanetary disk. Panel A depicts a time-series of the largest satellite embryo's physical radius. Facilitated by efficient gravitational focusing, the embryo experiences rapid initial growth. However, the rate of accretion slows down in time, as the velocity dispersion of the satellitesimal disk becomes progressively more excited. Intermittent jumps in radius correspond collisions of the proto-satellite with other massive embryos within the system. $84{,}000$ years into simulation time, the proto-satellite attains a radius of $2{,}600\,$km -- comparable to that of Ganymede. Panel B shows the cumulative size-frequency distribution of the system at $t=84{,}000\,$years. Importantly, this panel demonstrates that the aftermath of accretion within the circumplanetary disk is highly uneven, such that only two proto-satellites larger than $\mathcal{R}\geqslant1{,}000\,$km emerge at the end of the simulation. In addition to these two bodies, a single $\mathcal{R}\approx700\,$km object, along with seven $\mathcal{R}\approx500\,$km embryos occupy the $0.15-0.25\Rh$ satellitesimal annulus. On the smaller end of the size-frequency distribution, a prolonged tail of collisionally generated debris extends below $\mathcal{R}\lesssim0.1\,$km. Cumulatively, this calculation suggests that conditions within the circumplanetary disk are propitious to the emergence of a small number of massive embryos.}
\label{F:Boulder} 
\end{figure*}

In order to test the growth of a massive satellite embryo in the satellitesimal disk described in section \ref{sec:olig}, we have used a particle-in-a-box code \texttt{Boulder}, developed and described in \citet{Morbidelli2009}. The code accounts for the self-stirring of eccentricities and inclinations of the satellitesimal disk, as well as collisional damping, gas drag and dynamical friction. The latter damps the eccentricity and inclination of the most massive objects at the expense of causing the smallest particles to become dynamically excited. Collisions between particles were treated according to the prescription of \citet{1999Icar..142....5B} such that they could result in perfect merging, partial accretion, erosion, or catastrophic break-up, depending on collision velocities and sizes of the impacting bodies. 

Our initial conditions represent an annulus of debris centered at $r=0.2\,\Rh$ with a full width of $\Delta r=0.1\,\Rh$ i.e., the middle of the range illustrated in Figures (\ref{F:Taccr}-\ref{F:rF}), spanning $0.15$ to $0.25\,\Rh$. Initial satellitesimals were assigned a mass of $m=10^{19}\,$kg and a radius of $100\,$km in agreement with the estimate of section \ref{sec:tesimal}. The total mass of the satellitesimal population in the annulus was taken to be $M_{\rm{disk}}=3.2\times10^{23}\,$kg. This corresponds to a gas density of $\Sigma=1{,}000$\,g/cm$^2$ at $r=0.2\,\Rh$ in approximate agreement with our nominal profile (\ref{eqn:sigma}), a solid-to-gas ratio of $\mathcal{Z}=0.3$ and a satellitesimal formation efficiency of 30\%, uniformly spread over the annulus (the same parameters have been used in Figure \ref{F:rF}). Initially the eccentricities and inclinations were set to $\langle e \rangle \approx 6\times10^{-6}$ and $\langle i \rangle \approx 1.5\times10^{-5}\deg$, respectively. Importantly, however, these quantities evolved rapidly, such that after only $100$ years, the eccentricity and inclination $100\,$km satellitesimals were already $\langle e \rangle = 0.014$ and $\langle i \rangle=0.4\deg$ respectively, growing further to $\langle e \rangle =0.04$ and  $\langle i \rangle=1\deg$ by the $t=1{,}000\,$year mark.

The key advantage of a code like \texttt{Boulder} over the analytic calculations presented in section \ref{sec:olig} is that the code computes the gravitational focusing factor self-consistently, from the masses of the colliding bodies and their mutual velocity. That is, because of dynamical excitation within the disk, for a given target, the gravitational focusing factor decreases over time. However, because the most massive bodies grow more readily \citep{1972epcf.book.....S}, their focusing factor can instead increase, provided their escape velocity increases faster than the velocity dispersion in the disk.

Panel A of Figure (\ref{F:Boulder}) depicts the growth of the largest object within the annulus as a function of time. A radius of $\mathcal{R}=2{,}600\,$km is reached in a bit less than $100{,}000\,$years -- about an order of magnitude faster than estimated in Figure (\ref{F:Taccr}), where gravitational focusing is neglected.  We also note that embryo growth is more complex than that envisioned in section \ref{sec:olig} (i.e., accretion at a constant rate), as mergers with other massive bodies cause the satellite embryo's radius to sporadically jump upwards. Indeed, this is typical of the oligarchic growth process \citep{1998Icar..131..171K}.

Panel B of Figure (\ref{F:Boulder}) shows the cumulative size-frequency distribution of the satellitesimal population after $84{,}000\,$years. In addition to the aforementioned $\mathcal{R}=2{,}600\,$km body, there is a second body slightly exceeding $1{,}000\,$km in radius. Further down the radius ladder, there is one body with $\mathcal{R}\sim 700\,$km and seven with $\mathcal{R}\sim 500\,$km. In summary, only a handful of massive satellite embryos emerge form the annulus, with the vast majority of objects remaining small, or even decreasing in size because of collisional fragmentation.

\subsection{$N$-body Simulations} \label{sec:mercury}

\begin{figure*}[tbp]
\centering
\includegraphics[width=\textwidth]{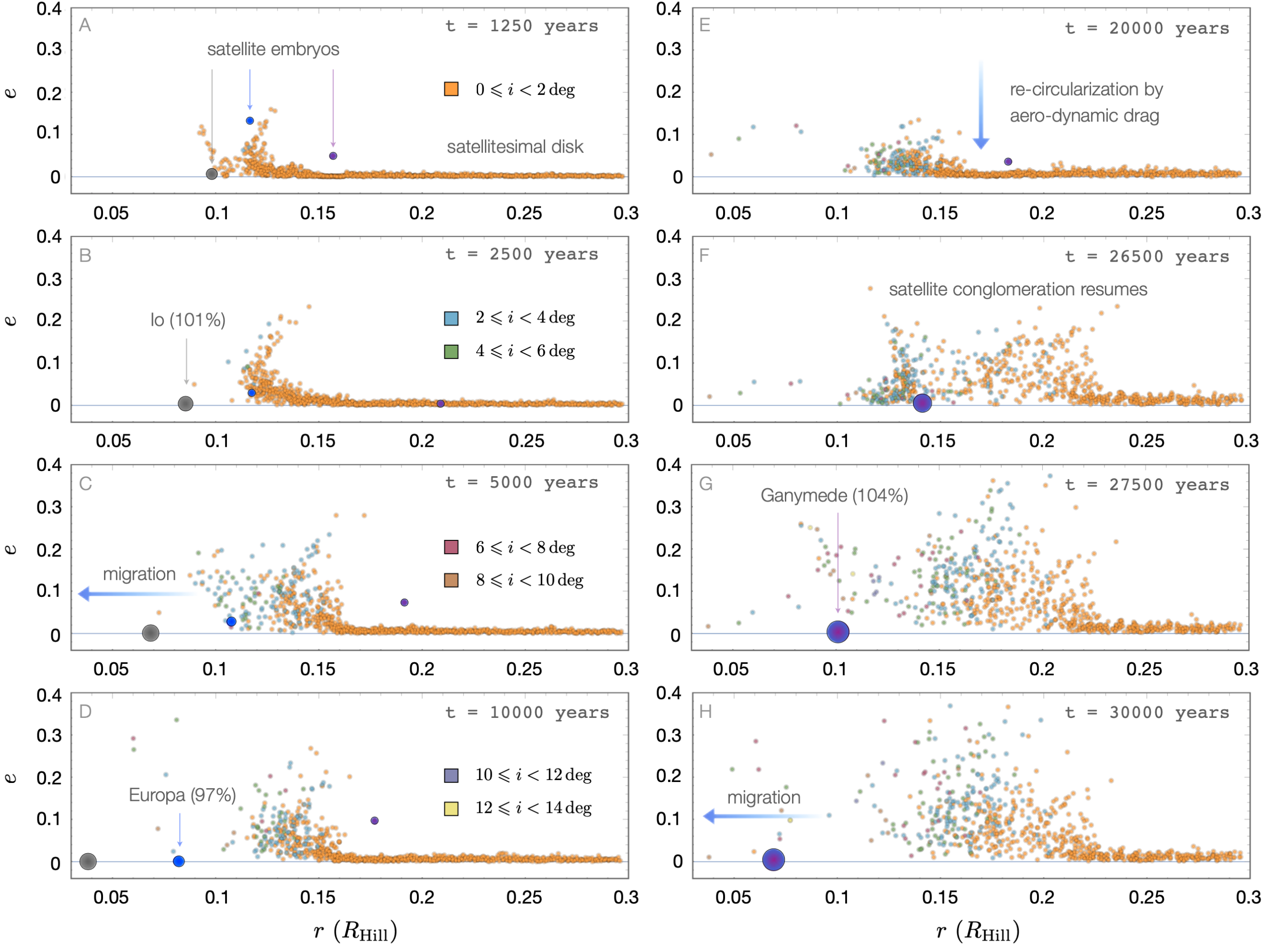}
\caption{Formation of the three inner Galilean satellites. An initially dynamically cold ($e\sim i\sim 10^{-3}$) disk of 1,000 super-satellitesimals, comprising $M_{\rm{disk}}=6\times10^{-4} \M$ is assumed to form by gravitational fragmentation between $r_{\rm{in}}=0.1\Rh$ and $r_{\rm{out}}=0.3\Rh$ (see section \ref{sec:tesimal}). Three satellite seeds are introduced within the same orbital range, at random planetocentric distances. In terms of disk metallicity, satellitesimal formation efficiency and the Safronov number, these initial conditions translate to $\mathcal{Z}\sim\Lambda\sim1/3$ and $\Theta\sim400$. Satellite seeds destined to become Io, Europa and Ganymede are depicted in gray, blue, and purple respectively, and their sizes serve as a proxy for their physical radii. On the other hand, colors of semi-active super-particles inform their orbital inclinations, as shown on the left column. In addition to gravitational dynamics, the effects of aerodynamic drag and gas disk-driven migration are self-consistently modeled in this simulation. \\
Results of this numerical experiment are summarized as follows. Owing to gravitational focusing in an initially pristine disk, conglomeration of Io begins quickly, and unfolds on a relatively short ($\sim1{,}000\,$year) timescale (panel A). $2{,}500\,$years into the simulation, Io decouples from the satellitesimal feeding zone and begins to migrate towards the Jovian magnetospheric cavity (panel B). As Io's orbit decays, Europa's growth ensues (panel C). However, due to an already-excited velocity dispersion among satellitesimals, Europa's accretion is somewhat less efficient, and by the $\sim10{,}000\,$year mark, Europa detaches from the satellitesimal disk, having achieved a smaller terminal mass than Io (panel D). For the following $\sim10^4\,$years, aerodynamic drag acts to re-cicularize the satellitesimal disk (panel E), and the conglomeration process restarts approximately $\sim25{,}000\,$years into the simulation (panel F). Ganymede achieves its terminal mass shortly thereafter (panel G), and by $30{,}000\,$years, follows Io and Europa on an inward migratory trek (panel H).}
\label{F:IEG} 
\end{figure*}

The above simulation demonstrates that the planetesimal sub-disk generated by gravitational collapse of dust is conducive to the emergence of isolated massive embryos. This particle-in-a-box calculation, however cannot capture the global dynamics of the system, which must instead be modeled with the aid of direct $N$-body simulations. Accordingly, we have carried out a series of numerical experiments that track the long-term orbital evolution of growing embryos, subject to gravitational coupling as well as disk-satellite interactions.

The initial conditions adopted in our $N$-body experiments draw upon the results of sections \ref{sec:dust}, \ref{sec:tesimal}, and \ref{sec:boulder}. In particular, our simulations began with a dynamically cold ($\langle e\rangle \sim \langle i \rangle \sim10^{-3}$) sea of satellitesimals, extending from $r_{\rm{in}}=0.1\Rh$ to $r_{\rm{out}}=0.3\Rh$. The effective disk surface density followed a $\propto r^{-5/4}$ profile as dictated by equation (\ref{eqn:powerlaw}), but with a diminished value of $\Sigma_0$. Keeping in mind that accretion is not expected to be 100\% efficient, the total mass of the planetesimal swarm was chosen to be $M_{\rm{disk}} = 6\times10^{-4}\,\M$ i.e., approximately three times the total mass of the Galilean satellites. We note that in terms of our model circumplanetary nebula outlined in section \ref{sec:disk}, this planetesimal disk is about an order of magnitude less massive than the cumulative gas mass contained in the same orbital region, and as before effectively translates to $\mathcal{Z}\sim \Lambda\sim 0.3$. 

To save computational costs, the planetesimal swarm was modeled as 1,000 semi-active $m\approx10^{21}\,$kg super-particles. These super-particles were allowed to gravitationally interact with the central planet and the satellite embryos, but not among themselves. Each super-satellitesimal was also subjected to aerodynamic drag ensuing from the circumplanetary nebula, employing the acceleration formulae of \citet{Adachi1976}. Despite being two orders of magnitude more massive than satellitesimals that are envisioned to result from gravitational fragmentation of the solid sub-disk, the aerodynamic drag calculation was carried out treating the particles as $\mathcal{R}=100\,$km, $\bar{\rho}=1\,$g/cc bodies.

The simulations were initialized with three satellite seeds (a separate discussion of the formation of Callisto will be presented below) with negligible masses, placed randomly between $0.1\,\Rh$ and $0.3\,\Rh$. Collisions between these proto-satellites and satellitesimals were treated as perfect mergers. In addition to conventional $N$-body interactions with the central planet and the planetesimal swarm, the satellite embryos experienced both aerodynamic drag (computed self-consistently, assuming $\bar{\rho}=1\,$g/cc), as well as type-I migration and orbital damping, which were implemented using the formulae of \citet{2000MNRAS.315..823P}. The migration and eccentricity/inclination damping timescales were taken to be $\mathcal{T}_{\rm{mig}}$ (equation \ref{eqn:Tmig}) and $\mathcal{T}_{\rm{damp}}=(h/r)^2\,\mathcal{T}_{\rm{mig}}=10^{-2}\,\mathcal{T}_{\rm{mig}}$ respectively. 

The calculations were carried out using the \texttt{mercury6} gravitational dynamics software package \citep{1999MNRAS.304..793C}. The hybrid Wisdom-Holman/Bulirsch-Stoer algorithm \citep{1991AJ....102.1528W,Press1992} was used throughout, with a time-step of $\Delta t=1\,$day and an accuracy parameter of $\hat{\epsilon}=10^{-8}$. Any objects that attained a radial distance in excess of a Jovian Hill radius were removed from the simulation. Additionally, any objects that attained an orbital radius smaller than $0.03\,\Rh$ (roughly the present-day semi-major axis of Ganymede) were absorbed into the central body. This was done to maintain a reasonably long time-step, with the understanding that the process of capturing Io, Europa and Ganymede into the Laplace resonance would have to be simulated a-posteriori.
 
We ran 12 such numerical experiments in total, each spanning $0.1\,$Myr. Qualitatively, simulation results followed the expectations of analytical theory outlined in the previous section. That is, growth of typical satellite embryos was terminated primarily by their departure from the debris disk through inward type-I migration. Moreover, the satellite conglomeration process -- once complete -- would leave behind a dynamically excited sea of satellitesimals, preventing the next satellite from forming until the system would re-circularize by aerodynamic drag.

Figure (\ref{F:IEG}) shows a series of snapshots of one particularly successful run, spanning $\sim1-30\,$kyr. This specific simulation yields three satellites that bear a striking resemblance to Io, Europa, and Ganymede both in terms of mass as well as orbital ordering. Within the context of this numerical experiment, owing to an initially low velocity dispersion among satellitesimals, the first (closest in) satellite seed experiences rapid growth (panel A). In only $\sim2\,$kyr, the satellite attains a mass equal to 101\% of Io, and sets off on an extended course of orbital decay (panel B). Meanwhile, a second embryo begins its conglomeration process (panel C). However, due to a pre-excited orbital distribution of satellitesimals, this embryo grows more slowly, and leaves the satellitesimal disk at the $\sim10\,$kyr mark, having attained a lower mass, equal to 97\% of Europa (panel D).

For $\sim10\,$kyr that follow, a large orbital eccentricity and inclination of the outermost satellite seed is maintained by the dynamically hot satellitesimal swarm (panel E). However, aerodynamic drag eventually re-circularizes the debris, and runaway growth of the final seed ensues approximately $25\,$kyr into the simulation (panel F). As with the first embryo, conglomeration proceeds rapidly, and the final seed reaches a mass equal to 104\% of Ganymede in only a few thousand years (panel G). By the $30\,$kyr mark, the third satellite leaves the satellitesimal feeding zone, and sets off on a steady path of inward migration.

\begin{figure}[tbp]
\centering
\includegraphics[width=\columnwidth]{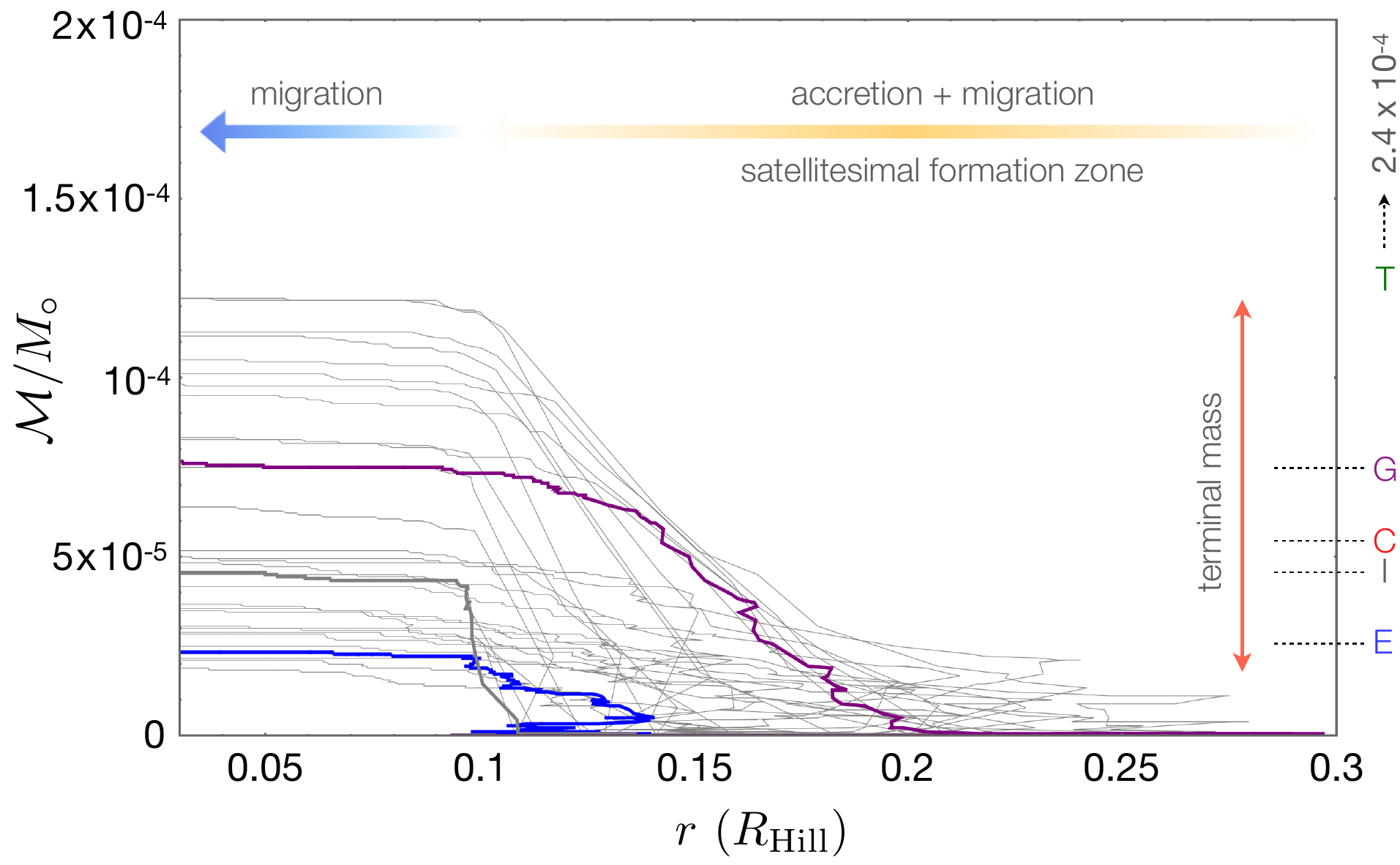}
\caption{Satellite formation tracks obtained within our full simulation suite, shown on a mass-orbital radius diagram. Generally, the terminal mass of objects generated in our $N$-body experiments is similar to that of the real Galilean satellites. Furthermore, the correct mass-ordering of the bodies is reproduced in 4 out of 12 instances. The specific formation tracks of Io, Europa and Callisto shown in Figure (\ref{F:IEG}) are highlighted with colored lines.}
\label{F:AM} 
\end{figure}

While this particular simulation provides the best match to the actual Galilean satellite masses, it is not anomalous within the broader context of our simulation suite. In particular, almost all of our runs generated satellites with masses that are comparable (within a factor of $\sim3$) to that of Io, and 4 out of 12 simulations ended with the correct mass ordering, wherein the least massive satellite is generated in between two more massive ones. Figure (\ref{F:AM}) shows the outcome of our complete simulation suite where satellite mass is plotted as a function of the orbital radius. Results of the particular simulation depicted in Figure (\ref{F:IEG}) are highlighted with thick lines. As an additional check on our calculations, we have carried out similar simulations using the \texttt{symba} integrator packafge \citep{2000AJ....120.2117L} employing marginally different implementation of aerodynamic drag and type-I migration, as well as a different $N$-body algorithm, and obtained similar results.

As a corollary, we remark that in some of our simulations, an inward-migrating Io captured a few satellitesimals into interior resonances, shepherding them onto very short-period orbits around Jupiter. Within the context of our model, we may envision that after the dissipation of the circumplanetary disk, a tidally receding Io would break resonance with these bodies, leaving them to encircle Jupiter to this day. Such a picture is remarkably consistent with the existence of Amalthea group of Jovian satellites -- a collection of four $\mathcal{R}\sim10-100\,$km objects possessing $P\sim7-16\,$hour orbital periods.
 
\subsection{Formation of the Laplace Resonance}
Among the most iconic and well-known characteristics of the three inner Galilean satellites is their multi-resonant orbital architecture. While an understanding of the celestial machinery of this resonance dates back to the work of Laplace himself, the dynamical origin of the 4:2:1 commensurability was only elucidated a little over half a century ago. In particular, \citet{Goldreich1965} was the first to propose that slow outward migration, facilitated by tidal dissipation within Jupiter, provides a natural avenue for the sequential establishment of a multi-resonant lock among the inner satellites. In the decades that followed, the plausibility of the tidal origin hypothesis was further corroborated with increasingly sophisticated numerical models \citep{1976ARA&A..14..215P,Henrard1982,2020arXiv200101106L}. 

An alternative picture -- proposed by \citet{PealeLee2002} -- is that although tidal dissipation is undoubtably an active process, the 4:2:1 orbital clockwork connecting Io, Europa, and Ganymede is primordial. More specifically, \citet{PealeLee2002} (see also \citealt{CanupWard2002}) suggest that the Laplace resonance was established before the dissipation of the circum-Jovian nebula as a result of convergent inward migration, driven by disk-satellite interactions. Because both the tidal migration and disk-driven migration scenarios can in principle reproduce the current orbital architecture of the satellites, it is difficult to definitively differentiate between them. Nevertheless, it is obvious that disk-driven assembly of the Laplace resonance ensues naturally within the context of our model, and to complete the qualitative narrative proposed herein, we explore this process numerically.

Recall that the $N$-body simulations carried out in the previous sub-section (and illustrated in Figure \ref{F:IEG}) point to sequential satellite formation, where upon accruing a sufficiently large mass, a growing object exits the satellitesimal disk via inward type-I migration. If the gaseous component of the circumplanetary disk were to extend down to the planetary surface, the in-spiraling satellite would simply be engulfed by the planet. However, as already mentioned in section \ref{sec:disk}, rudimentary considerations of the relationship between magnetic field generation and giant planet luminosity during final stages of accretion suggest that the circumplanetary disk is likely to be truncated by the planetary magnetosphere at a radius of $\Rt\sim5\,\Rjup$ \citep{Batygin2018,2020MNRAS.491L..34G}. Accordingly, the inner edge of the nebula should act as a trap that halts the orbital decay of the first large satellite (Io) at $r\approx\Rt$. 

We begin our simulations of Laplace resonance assembly at this stage. Io is assumed to start at its current orbital location (a value approximately equal to $\Rt$), while Europa and Ganymede are initialized out of resonance, with semi-major axes a factor of 2 and 4 greater than that of Io, respectively. Rather than attempting to emulate the effects of the satellite trap on Io through sophisticated parameterization of type-I migration (see e.g., \citealt{Izidoro2019}), here we opt for a simpler procedure wherein the semi-major axes of all satellites in the calculation are renormalized at every time-step\footnote{To carry out the simulations of Laplace resonance assembly, we employed the conventional Bulisch-Stoer algorithm, with an initial time-step of $\Delta t = 0.01\,$days.} such that the orbital period of the innermost body is always equal to that of Io (see e.g., \citealt{Deck2015} for more discussion). Convergent orbital evolution is simulated by applying the type-I migration torque \citep{2000MNRAS.315..823P} to Europa and Ganymede. For definitiveness, we adopt a common characteristic migration timescale for both objects, which maintains their non-resonant period ratio prior to Europa and Io's encounter with the 2:1 commensurability. At the same time, type-I eccentricity and inclination damping -- assumed to operate on a timescale a factor of $(h/r)^{-2}=100$ times shorter than the migration time \citep{Tanaka2004} -- is applied to all satellites.

\begin{figure}[tbp]
\centering
\includegraphics[width=\columnwidth]{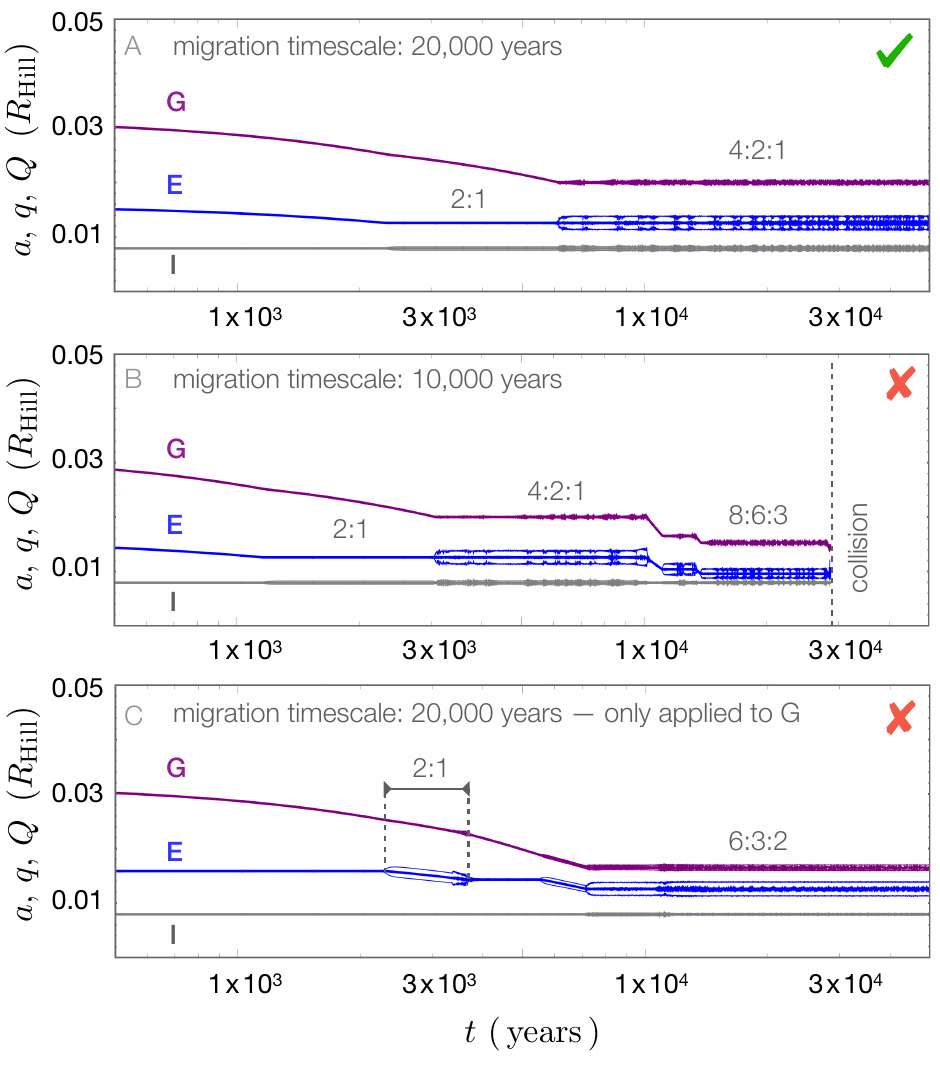}
\caption{Formation of the Laplace resonance. Panel A: migration of Europa and Ganymede towards Io on a $\mathcal{T}_{\rm{mig}}=20{,}000\,$year timescale. In this simulation, convergent orbital evolution of the three inner Galilean satellites leads to sequential locking of Io, Europa, and Ganymede into a long-term stable 4:2:1 mean-motion commensurability. This sequence of events is consistent with the actual architecture of Jovian satellites. Panel B: if the convergent migration timescale is reduced by a factor of two (such that $\mathcal{T}_{\rm{mig}}=10{,}000\,$years), the 4:2:1 Laplace resonance is rendered long-term unstable. In this case, after the satellites break out of the 4:2:1 commensurately, they temporarily get captured into a more compact 8:6:3 resonance. However, this configuration is also long-term unstable, and eventually a full-fledged orbital instability develops, triggering satellite collisions. Panel C: a demonstration of resonant over-stability in the Galilean system. If Europa and Ganymede lock into the 2:1 commensurability before Io and Europa do, the associated dynamics are over-stable, and in due course, the full system equilibrates within the 6:3:2 -- rather than the 4:2:1 -- multi-resonant configuration. Cumulatively, these numerical experiments point to two independent constraints. First, if the Laplace resonance is primordial, Io and Europa must have locked into the 2:1 resonance before Europa and Ganymede approached a 2:1 commensurability. Second, the timescale for convergent migration could not have been much shorter than $\sim20{,}000\,$years.
}
\label{F:LapRes} 
\end{figure}

Figure (\ref{F:LapRes}A) shows the results of our fiducial numerical experiment, where the migration timescale is set to $\mathcal{T}_{\rm{mig}}=20,000\,$years. Qualitatively, the satellites follow the same evolutionary sequence as that outlined in the simulations of \citet{PealeLee2002}. Namely, Europa reaches the 2:1 commensurability with Io first, leading to the establishment of a resonant lock. An interplay between resonant dynamics and continued type-I torque exerted on Europa adiabatically excites the eccentricities of both inner satellites, until this process is stabilized by disk-driven eccentricity damping. Eventually, Ganymede reaches a 2:1 resonance with Europa, and a long-term stable 4:2:1 multi-resonant chain is established. As already pointed out in the work of \citet{PealeLee2002}, the resonance established through convergent migration within the circumplanetary nebula is a different variant of the Laplace resonance than the one the satellites occupy today (meaning that the Laplace angle exhibits asymmetric libration with an appreciable amplitude instead of being tightly confined to $180\,$deg). However, their calculations also demonstrate that as soon as satellite-disk interactions subside and are replaced with conventional tidal evolution, the satellites' eccentricities rapidly decay, leading to the establishment of the observed resonant architecture.

\paragraph{The Adiabatic Limit} We note that the migration timescale adopted in our fiducial numerical experiment exceeds the theoretical value given by equation (\ref{eqn:Tmig}) by a factor of $\sim 5$. We do not consider this to be a meaningful drawback of our model because it is unlikely that our simplified description of the circumplanetary disk can predict the actual migration rate experienced by the Galilean satellites to better than an order of magnitude. An arguably more important consideration is that independent of any particular formation scenario, the masses of the satellites dictate a minimum orbital convergence timescale, below which the establishment of a long-term stable 4:2:1 mean motion commensurability becomes improbable. Simply put, this is because adiabatic capture into a mean motion resonance requires the resonance bandwidth crossing time to significantly exceed the libration timescale of the resonant angles. Notably, the former is set by the assumed migration timescale while the latter is determined by the satellite masses (see e.g., \citealt{Batygin2015} and the references therein).

To quantify the adiabatic limit of the rate of orbital convergence among Jovian satellites, we repeated the aforementioned experiment, reducing the migration timescale by a factor of two, such that $\mathcal{T}_{\rm{mig}}=10,000\,$years. The corresponding results are depicted in the middle panel (B) of Figure (\ref{F:LapRes}). Although the early stages of this simulation resemble the evolution depicted in Figure (\ref{F:LapRes}A) (in that the satellites do get temporarily locked into a 4:2:1 resonance), in a matter of a few thousand years, they break out of this configuration and following a transient period of chaotic dynamics, stabilize in a more compact 8:6:3 resonant chain. Even this configuration, however, is not immutable: approximately 30,000 years into the simulation, the system becomes dynamically unstable, and collisions among satellites ensue shortly thereafter. Thus, we conclude that if the Laplace resonance is indeed primordial, migration timescale associated with the orbital assembly of Galilean satellites could not have been much shorter\footnote{Importantly, independent of the details of the accretion process, this requirement for a relatively long migration timescale necessitates a low-mass circumplanetary disk.} than 20,000 years.

\paragraph{Over-stability} Apart from the migration timescale itself, a separate constraint on the assembly of the Laplace resonance concerns the order in which the observed orbital architecture was established. Recall that within the context of the simulations described above, Europa encountered the 2:1 mean motion commensurability with Io \textit{before} Ganymede joined the resonant chain. This is due to the fact that within the framework of our model, satellite formation is envisioned to occur successively rather than simultaneously. Indeed, had all three satellites emerged within the circumplanetary disk at the same time, the Europa-Ganymede resonance would have been established first, since Ganymede is more massive and would have experienced more rapid orbital decay. As it turns out, sequential formation of satellites is not simply a natural outcome of our theoretical picture (as depicted in Figure \ref{F:IEG}) -- it is a veritable requirement of the observed resonant dynamics.

An intriguing aspect of disk-driven resonant encounters is that the long-term stability of the ensuing resonance can be compromised by the same dissipation that leads to its establishment. This effect -- known as resonant over-stability -- exhibits a strong dependence on the satellite mass ratio and manifests in systems where the outer secondary body is more massive than the inner \citep{Goldreich2014,Deck2015,Xu2018}. To this end, the analytic criterion for over-stability (see Figure 3 of \citealt{Deck2015}) suggests that the factor of $\sim 3$ difference between the masses of Ganymede and Europa is sufficient to render the 2:1 resonance unstable, if the satellite pair encounters it in isolation. In other words, over-stability of resonant dynamics indicates that the observed 4:2:1 Laplace resonance could not have been established if Ganymede and Europa locked into resonance before Europa and Io did.

To confirm this anticipation, we repeated the above numerical experiment, restoring $\mathcal{T}_{\rm{mig}}$ to $20,000\,$years but only applying the migration torque to Ganymede (conversely, eccentricity damping torque was applied to all satellites as before). This choice ensured that Ganymede would encounter the 2:1 resonance with Europa first, since in this experiment Europa experiences no explicit disk-driven migration. The results of this simulation -- depicted in Figure (\ref{F:LapRes}C) -- followed analytic expectations precisely: upon entering the 2:1 resonance, over-stable librations ensued, propelling Europa and Ganymede to break out of the 2:1 resonance before the establishment of the 4:2:1 resonant chain. Eventually, convergent migration did drive the system into a multi-resonant configuration, but it was characterized by a 6:3:2 period ratio. Indeed, the Io-Europa-Ganymede Laplace resonance appears to have been built from the inside out.

\subsection{A Final Wave of Accretion}

Up until this point, we were primarily concerned with the conglomeration and migration of the inner three Galilean satellites within the gaseous circumplanetary disk. But what happens when the photo-evaporation front reaches the giant planets' orbits and the gas is removed? One trivial consequence of gas removal is that the metallicity of the system $\mathcal{Z}\rightarrow \infty$ everywhere in the disk. Referring back to equation (\ref{eqn:Q}), this would imply that the $Q_{\bullet}\lesssim 1$ condition would be satisfied at all orbital radii (i.e., not just the outer disk as shown in Figure \ref{F:Q}), implying the onset of a final wave of satellitesimal formation. Accordingly, let us now consider the growth of a satellite embryo within this gas-free environment, with an eye towards quantifying the formation of Callisto and Titan.

\begin{figure*}[tbp]
\centering
\includegraphics[width=\textwidth]{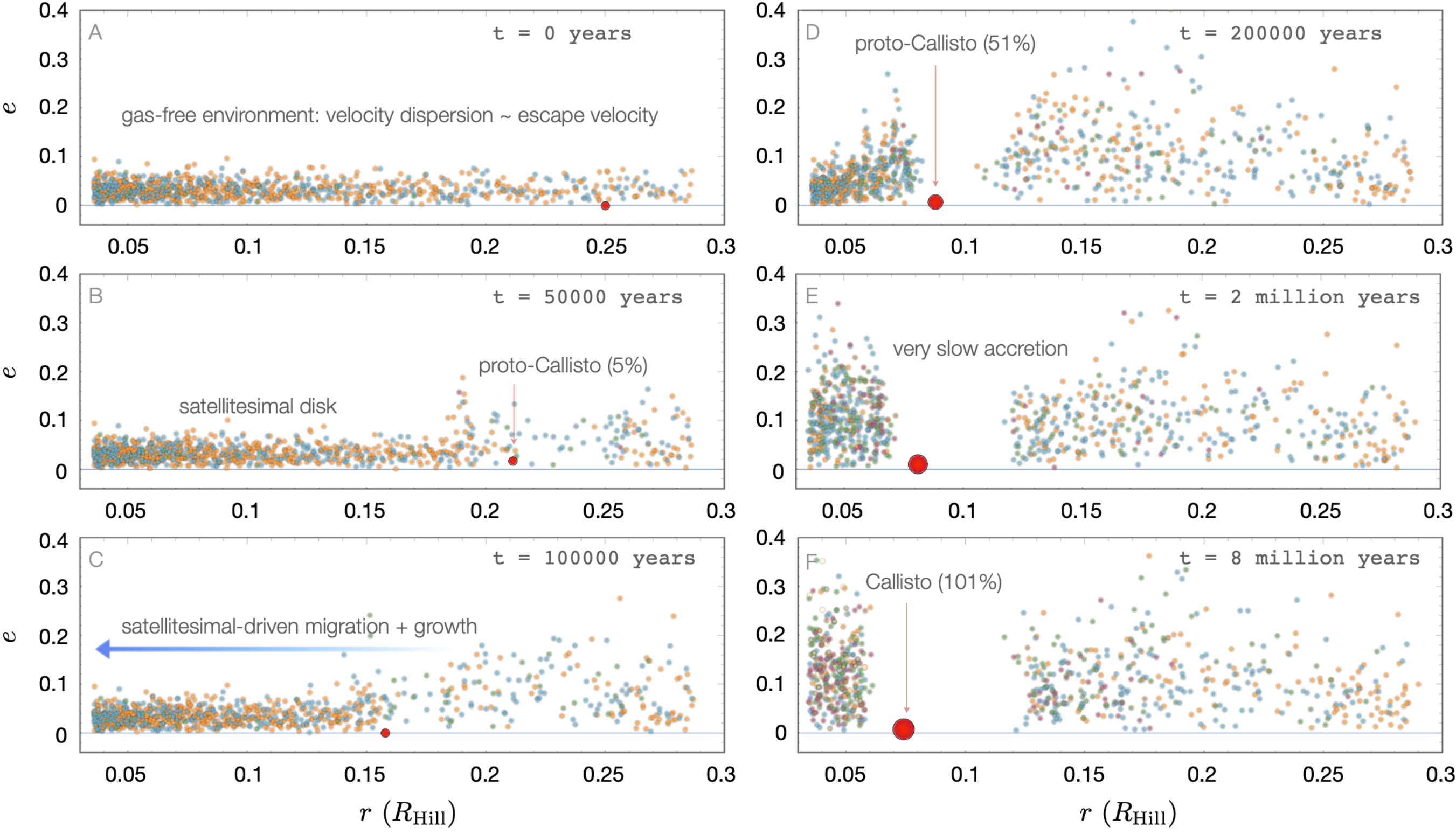}
\caption{Formation of Callisto in a gas-free satellitesimal swarm. A similar accretion scenario can be envisioned for Titan. A disk of debris, comprising $M_{\rm{disk}}=2\times10^{-4}\,\M$ was initialized between $r_{\rm{in}}=0.03\,\Rh$ and $r_{\rm{out}}=0.3\,\Rh$, with a Safronov number of order unity: $\Theta\sim 1$ (panel A). A single satellite seed -- depicted in red -- is introduced at $r=0.25\,\Rh$. Unlike the results reported in Figure (\ref{F:IEG}), in absence of dissipative effects associated with the presence of the circumplanetary disk, the initial phase of satellite conglomeration proceeds slowly. Correspondingly, 50,000 years into the simulation, the embryo has only acquired $5\%$ of Callisto's mass (panel B). In the following 100,000 years, growth temporarily accelerates, and in concert with the accretion process the satellite embryo migrates inward by scattering satellitesimals (panel C). By the 200,000 year mark, the satellite embryo is a factor of two less massive than Callisto, but due to orbital excitation of neighboring debris, satellitesimal-driven migration effectively grinds to a halt (panel D). Subsequently, the rate of accretion slows down dramatically, such that 2 million years after the start of the simulation, proto-Callisto has only reached about three quarters of its terminal mass (panel E). The embryo finally achieves Callisto's actual mass after 8 million years of evolution (panel F).}
\label{F:Callisto} 
\end{figure*}

In terms of basic characteristics, Callisto and Titan share many similarities. Both have orbital periods slightly in excess of two weeks (corresponding to approximately $3.5\%$ and $2\%$ of Jupiter and Saturn's respective Hill spheres). The physical radii and masses of the satellites are also nearly identical (although when normalized by the masses of their host planets, Titan is larger than Callisto by a factor of $\sim4$). Finally, measurements of the satellites' axial moments of inertia through spacecraft gravity data point to the distinct possibility that these satellites may be only partially differentiated (\citealt{Anderson2001,Iess2010}; see however \citealt{GaoStevenson2013}), implying a formation timescale that exceeds $\sim0.5\,$Myr \citep{BarrCanup2008}. Accordingly, let us now examine if the post-nebular phase of our model can naturally generate bodies sharing some of Callisto and Titan's attributes. For definitiveness, in the remainder of this section we will concentrate on the formation of Callisto, although the applicability of the calculations to Titan is also implied.

We begin by envisioning proto-Callisto as a satellite embryo embedded in a disk of icy debris. When submerged in a sea of solid material, the embryo interacts with its environment by gravitationally stirring the satellitesimal swarm \citep{1972epcf.book.....S}. This process has two direct consequences: collisions with small bodies lead to steady accretion, while asymmetric scattering of debris facilitates transfer of angular momentum \citep{Ida2000,Kirsh2009}. Therefore, in a gas-free environment, growth of the embryo must occur concurrently with satellitesimal-driven migration. Although some analytic understanding of the associated physics exists \citep{2014Icar..232..118M}, $N$-body simulations provide the clearest illustration of the ensuing dynamical evolution. Correspondingly, in an effort to maintain a closer link with other calculations presented in this section, we continue on with a numerical approach.

Our simulation setup essentially constituted a stripped-down, purely gravitational variant of the calculations presented in section \ref{sec:mercury}. More specifically, a disk of 1,000 super-satellitesimals comprising $M_{\rm{disk}} = 2\times10^{-4}\,\M$ was initialized between $r_{\rm{in}}=0.03\Rh$ and $r_{\rm{out}}=0.3\Rh$, together with a single satellite embryo residing in the outer disk at $r=0.25\Rh$. This swarm of satellitesimals is envisioned to have coalesced from the remainders of debris left behind in the aftermath of the accretion of Io, Europa, and Ganymede, as well as new satellitesimals, formed from the left-over dust in a wave of gravitational collapse triggered by the dissipation of the gas. Owing to the envisioned lack of residual Hydrogen and Helium within the system, both the effects of aerodynamic drag as well as type-I orbital migration were assumed to be negligible. Moreover, with no gas to dynamically cool the system, we assumed that the initial velocity dispersion of the satellitesimals was set by gravitational self-stirring, and was therefore comparable to the escape velocity of the small bodies $\langle v \rangle \sim v_{\rm{esc}}$. All other details of the numerical calculations were identical to those reported in \ref{sec:mercury}.

As before, 12 numerical experiments were carried out, but with the integration time increased to $10\,$Myr. We note that the assumed inner edge of the planetesimal disk lies slightly interior to Callisto's current orbital semi-major axis, and approximately coincides with the location of Ganymede's exterior 2:1 resonance. In light of this correspondence, it is natural to expect that as proto-Callisto approaches its final orbit, satellitesimals at the inner edge of the particle disk will experience a complex interplay of perturbations arising from Callisto itself as well as from Ganymede. Nevertheless, to subdue the already-formidable computational costs, and maintain a reasonably long time-step, we disregard the existence of the inner three Galilean satellites in these simulations. Although this assumption does not pose a significant problem throughout most of the simulation domain, the dynamical evolution exhibited by the system close to $r_{\rm{in}}$ should indeed be viewed as being highly approximate.

A variant of Figure (\ref{F:IEG}) pertinent to the formation of Callisto is shown in Figure (\ref{F:Callisto}). Initial conditions characterized by a Safronov number of order unity are depicted in Panel A. Although embryo growth accompanied by gravitational stirring of the disk starts immediately (panel B), it proceeds very slowly at first. A comparatively rapid phase of accretion and satellitesimal-driven migration ensues $100{,}000$ years into the simulation (panel C), but terminates at $200{,}000$ years, with Callisto having achieved approximately half of its mass (panel D). Owing to an excited orbital distribution, subsequent growth unfolds on an exceptionally long timescale (panel E), such that Callisto only achieves its full mass $8\,$Myr into the simulation (panel F). 

We note that if left unperturbed, the remaining satellitesimal swarm beyond $r\gtrsim0.1\,\Rh$ would eventually coalesce into additional satellites. However, \citet{Deienno2014,Nesvorny2014} have demonstrated that any objects beyond the orbits of Callisto and Iapetus are readily destabilized by planetary flybys that transpire during the solar system's transient phase of dynamical instability. All satellites interior to Titan, on the other hand, can in principle be accounted for by the ring-spreading model of \citet{2012CridaCharnozMoonForm} (see also the work of \citealt{CharnozRingsSpreading}). Consequently, the radial extent of regular satellite systems of Jupiter and Saturn likely reflect a combination of processes including gravitationally focused pair-wise accretion, orbital migration, and external dynamical sculpting. 

\section{Discussion} \label{sec:disc}

%In parallel with theoretical studies of XXXXX, the recent years have seen renewed interest in icy satellites as physical bodies. Recent observational efforts have delineated evidence for XXXX. Detailed characterization are poised to intensify in the coming years, with NASA's Europa Clipper and ESA's JUICE missions headed for the Gallilean moons, as well as NASA's DragonFly mission headed for Titan. As the physical nature of these worlds comes into sharper focus, ... 

%We developed a new picture. Let us outline what this picture actually entails, at a qualitative level. 

Although subject to nearly-continuous astronomical monitoring for centuries, the natural satellites of Jupiter and Saturn have only come into sharper focus within the last forty years. The unprecedented level of detail unveiled by the Voyager flybys \citep{1979Sci...204..951S,1979Sci...206..927S} as well as the Galileo/Cassini orbiters \citep{Greeley1998,1998Icar..135..127M,1998Icar..135..276P,Elachi2005,2007Natur.445...61S,2016AREPS..44...57H} has incited a veritable revolution in our understanding of these faraway moons, once and for all transforming them from celestial curiosities into bonafide extraterrestrial worlds. This ongoing paradigm shift -- sparked during the latter half of the last century -- is poised to persist in the coming decades, as Europa Clipper, JUICE, and Dragonfly missions\footnote{JUICE, Europa Clipper, and Dragonfly missions are expected to launch in 2022, 2025, and 2026, respectively.}, along with ground-based photometric/spectroscopic observations \citep{2017AJ....153..250T,2019SciA....5.7123T,2019AJ....158...29D,2019GeoRL..46.6327D}, continue to deepen our insight into their geophysical structure. Importantly, all of these developments have added a heightened element of intrigue to the unfaltering quest to unravel the origins of natural satellites within the solar system \citep{1999ARA&A..37..533P}.

In parallel with \textit{in-situ} exploration of the sun's planetary album, detailed characterization of gas flow within young extrasolar nebulae \citep{2019ApJ...879L..25I,Teague2019} has began to illuminate the intricate physical processes that operate concurrently with the final stages of giant planet accretion. Coupled with high-resolution hydrodynamical simulations of fluid circulation within planetary Hill spheres \citep{Tanigawa2012,Morbidelli2014,Judit2014,Judit2016,Lambrechts2019}, these results have painted an updated portrait of the formation and evolution of circumplanetary disks. While this newly outlined picture has been instrumental to the successful interpretation of modern observations, it has also brought to light a series of puzzles that remain elusive within the context of the standard model of satellite formation \citep{CanupWard2002,CanupWard2006}. In particular, the physical process that underlies the agglomeration of sufficiently large quantities of dust within the circumplanetary disks, the mechanism for conversion of this dust into satellite building blocks, and primary mode by which satellitesimals accrete into full-fledged satellites have remained imperfectly understood.

In this paper, we have presented our attempt at answering these questions from first principles. Let us briefly summarize our proposed scenario. 

\subsection{Key Results}

Inspired by the aforementioned observational and computational results, we have considered the conglomeration of satellites within a vertically-fed, steady-state H/He decretion disk that encircles a young giant planet. Owing to pronounced pressure-support, gas circulation within this disk is notably sub-Keplerian, and is accompanied by a (viscously-driven) radial outflow in the mid-plane. Although the system originates strongly depleted in heavy elements, its metallicity is envisaged to increase steadily in time. More specifically, our calculations show that $s_{\bullet}\lesssim10\,$mm dust grains are readily trapped within our model circumplanetary disk, thanks to a hydrodynamic equilibrium that ensues from a balance between energy gains and losses associated with the radial updraft and azimuthal headwind, respectively. While it may be impossible to definitively prove that this process truly operated in gaseous nebulae that encircled Jupiter and Saturn during the solar system's infancy, our theoretical picture exhibits a remarkable degree of consistency with the recent observations of \citet{2019ApJ...884L..41B}, which demonstrate that the circumplanetary disk in orbit of the young giant planet PDS\,70c is enriched in dust by more than an order of magnitude compared with the expected baseline metallicity.

As the dust-to-gas ratio of the system grows, the solid sub-disk progressively settles towards the mid-plane, eventually becoming thin enough for gravitational fragmentation to ensue \citep{GoldreichWard1973}. Correspondingly, large-scale gravitational collapse of the dust layer generates a satellitesimal disk containing $M_{\rm{disk}}\sim6\times10^{-4}\,\M$ worth of material between $\sim0.1\,\Rh$ and $\sim0.3\,\Rh$. The velocity dispersion of the resulting satellitesimal swarm is heavily damped by aerodynamic drag originating from the gas, allowing for efficient capture of satellitesimals by an emerging satellite embryo \citep{1972epcf.book.....S,Adachi1976}. Assisted by gravitational focusing, this embryo continues accreting until it becomes massive enough to raise significant wakes within its parent nebula. The gravitational back-reaction of the spiral density waves upon the satellite results in orbital decay (e.g., \citealt{Tanaka2002}), terminating further growth of the newly formed satellite by removing it from the satellitesimal feeding zone. Eventually, the satellite reaches the inner boundary of the disk, and halts its migratory trek.% (the case of Jupiter). Alternatively, if the planetary magnetic field is insufficiently strong for the truncation radius to exceed the Roche radius, the satellite gets tidally disrupted and accreted by its host planet %(the case of Saturn).

An important attribute of the above picture is that as the satellite exists the feeding zone by orbital migration, it leaves behind a dynamically excited orbital distribution of satellitesimals. With nothing to facilitate continued gravitational stirring, however, satellitesimal orbits re-circularize and collapse back down to the equatorial plane under the action of aerodynamic drag. The satellite formation process then restarts, generating a second embryo. Eventually, this embryo also grows to its terminal radius, dictated by a near-equality of the migration and mass-doubling timescales (see also \citealt{CanupWard2006}), and subsequently also exits the satellitesimal swarm. If the proceeding satellite is retained within the inner region of the disk, convergent migration of the bodies facilitates locking into a mean-motion resonance. If this sequence of events occurs more than twice, a resonant chain -- akin to that exhibited by the three inner Galilean moons -- can be naturally generated \citep{PealeLee2002}. Our calculations further demonstrate that due to constraints associated with resonant over-stability \citep{Goldreich2014,Deck2015}, the Laplace resonance must have been assembled from the inside-out, and that the timescale for orbital convergence must have exceeded $\mathcal{T}_{\rm{mig}}\gtrsim20{,}000\,$years.

In principle, we can imagine that the process of sequential satellite generation continues until the gas is abruptly removed by photo-evaporation of the circumstellar nebula \citep{2012MNRAS.422.1880O}. To this end, we note that the gravitational binding energy of a Hydrogen molecule in orbit around Jupiter at $r\sim0.01\Rh$ is approximately equal to that of a Hydrogen molecule in orbit of the sun at $r\sim5\,$AU. This means that the same solar photons that successfully eject gas from the sun's potential well at Jupiter's orbit can also expel gas from the Jovian potential well at a distance comparable to the truncation radius. In turn, this implies that the removal of the gaseous components of the circumplanetary nebula and the circumstellar disk must occur simultaneously. 

Driven by a sharp increase in effective disk metallicity, the remaining dust sub-disk fragments into satellitesimals, setting the stage for the accretion of the final satellite embryo. However, unlike the comparatively rapid mode of satellite formation described above, in this gas-free environment, embryo growth proceeds on a multi-million year timescale, leading to only partial differentiation of the resulting body \citep{BarrCanup2008}. The mechanism of inward migration is also distinct in that it is facilitated by asymmetric scattering of debris rather than tidal interactions with the gas \citep{Kirsh2009}. For the specific purposes of this study, we consider the slow conglomeration process to be relevant to the formation of Callisto (and perhaps, Titan). As a concluding step to the narrative, we invoke the effects of planet-planet scattering during the transient phase of giant planet instability to disperse the remaining satellitesimal disk, leaving only the deepest segments of the planetary Hill spheres to host large natural satellites \citep{Deienno2014,Nesvorny2014}.

\subsection{Jupiter vs. Saturn}

Despite sharing some basic properties, the satellite systems of Jupiter and Saturn are far from identical, and it is worthwhile to contemplate how the differences between them came to be. In section \ref{sec:sims}, we asserted that the formation narratives of Callisto and Titan may be similar, leaving open the question of why Saturn does not possess an equivalent system of large, multi-resonant (Galilean-type) inner moons. Within the framework of our model, two separate explanations for this disparity can be conjured up. Perhaps most simply, we can envision a scenario where Saturn's circumplanetary nebula never achieved the requisite metallicity for large-scale fragmentation of the solid sub-disk (after all, agglomeration an overall metallicity in excess of $\mathcal{Z}\gtrsim0.1$ is not guaranteed). This would have delayed the process of satellitesimal formation until the dissipation of the gas, bringing the initial conditions of the Saturnian system in line with those assumed in section \ref{sec:sims}.

Alternatively, we may attribute the difference between Jovian and Kronian systems to a disparity in the host planets' ancient dynamos. That is, if Saturn's magnetosphere was insufficiently prominent to truncate the circumplanetary disk outside of the Roche radius, any inward-migrating satellites would have been tidally disrupted, leaving Titan as ``the last of the Mohicans" \citep{CanupWard2006}. We will resist the urge to speculate as to which of these imagined solutions may be more likely, and simply limit ourselves to pointing out that while the latter scenario would imply a primordial origin for Saturn's rings \citep{Canup2010TitanMakesRings,Crida2019SatRingsYoung}, the former picture is more consistent with the recently proposed ``young rings" hypothesis\footnote{While plausible, a qualitative mechanism that could feasibly generate the rings within the last few tens of Myr remains elusive.} \citep{2013Icar..223..544A,Cuk2016SatRingsYoung,Dubinski2019RecentFormationofRings}.

\subsection{Criticisms \& Future Directions}

Although the calculations summarized above outline a sequential narrative for the formation of giant planet satellites, much additional work remains to be done before our model can be considered complete on a detailed level. Accordingly, let us now propose a series of criticisms of the envisioned scenario, and delineate some avenues for future development of the theoretical picture.

Arguably the most basic critique of our model concerns a coincidence of timescales. More specifically, we have imagined that satellite formation - despite requiring hundreds of thousands of years to complete - unfolds in a steady-state circumplanetary disk that encircles an already-assembled giant planet. In order for this picture to hold, two criteria have to be satisfied. First, the appearance of the circumplanetary disk must concur with the concluding epochs of Jupiter and Saturn's respective phases of rapid gas-agglomeration (although some theoretical evidence that supports this notion already exists within the literature, additional work is undoubtably required; \citealt{Judit2016,Lambrechts2019}). Second, after concluding their formation sequences, the solar system's giant planets would have had to reside inside the protosolar nebula for an extended period of time without experiencing appreciable additional growth\footnote{Notably, such a sequence of events is required for the so-called Grand Tack scenario of early solar system evolution \citep{2011Natur.475..206W}.}. This would imply the existence of a yet-to-be-characterized process that suspends runaway accretion of a giant planet within a long-lived circumstellar nebula. Although a physical mechanism that could robustly regulate the terminal masses of giant planets remains elusive \citep{Morbidelli2014,GinzburgChiang2019}, the emergence of an extended planetary magnetosphere due to an enhanced surface luminosity presents one possible option that merits further consideration \citep{Batygin2018,Cridland2018}.

A distinct issue relates to the dust-gas equilibrium derived in section \ref{sec:dust}. In particular, the grain radii encapsulated by the headwind-updraft balance outlined in this work are relatively limited (i.e., $s_{\bullet} \sim 0.1-10\,$mm), and this restriction illuminates grain-growth as a potential pathway for dust particles to break out of equilibrium. While the cycle of nucleation followed by sublimation of volatile species inside the circumplanetary disk's ice-line discussed in section \ref{sec:dust} should modulate the grain radii to remain within the aforementioned range, the viability of this mechanism remains to be demonstrated quantitatively. Finally, the process of satellitesimal formation within the circumplanetary disk deserves more careful scrutiny. That is, although energy-limited settling of dust followed by gravitational fragmentation invoked in section \ref{sec:tesimal} provides a particularly simple scenario for conversion of dust into satellite building blocks, it is entirely plausible that a detailed examination of high-metallicity dust-gas dynamics will reveal a more exotic mode of satellitesimal formation that has eluded our analysis.

\subsection{Planet vs. Satellite Formation}
We began this paper by highlighting a similarity between detected systems of extrasolar Super-Earths and the moons of giant planets. Correspondingly, let us conclude this work with a brief comment on the relationship between the machinery that underlies our proposed theory of satellite accretion, and the standard theory of planet formation \citep{Armitage2010}. Undeniably, some analogies must exist between our model and the standard narrative of Super-Earth conglomeration, since inward orbital migration, which is terminated by a steep reduction in the gas surface density due to magnetic truncation of the disk (e.g., \citealt{Masset2006,Izidoro2019}) plays a notable role in both cases. The parallels, however, may end there. 

Recall that within the context of our picture, satellite building blocks are envisioned to form via the Goldreich-Ward mechanism, which is in turn facilitated by a hydrodynamic equilibrium that can only be achieved in a decretion disk. This view is in stark contrast with the now-widely accepted model of planetesimal formation, which invokes the streaming instability, (although both processes culminate in gravitational collapse and generate $\mathcal{R}\sim100\,$km objects; \citealt{YoudinGoodman2005,JohansenYoudin2007}). Moreover, pebble accretion, which is increasingly believed to drive the conglomeration of giant planet cores and Super-Earths alike \citep{LambrechtsJohansen2012,2015A&A...582A.112B} is unlikely to play the leading role in facilitating satellitesimal growth. Instead, satellite accretion proceeds via pair-wise collisions among satellitesimals, entailing a closer link to the physics of terrestrial planet formation \citep{Lissauer1993,2009ApJ...703.1131H,2016AJ....152...68W,2018A&A...612L...5O} than anything else. Cumulatively, these contrarieties suggest that the model outlined herein cannot be readily applied to circumstellar disks, and the architectural similarities between solar system satellites and short-period extrasolar planets are likely to be illusory.

%we note that once the photo-evaporation front reaches the orbit of the host planet and removes the surrounding gas, the gaseous component circumplanetary disk cannot be retained either. This is evident from a simple energy argument: the ratio the gravitational binding energy of a Hydrogen molecule in orbit around Jupiter at $r\sim0.01\Rh$ is approximately equal to that of a Hydrogen molecule in orbit of the sun at $r\sim5\,$AU. This means that the same solar photons that successfully eject gas from the sun's potential well at Jupiter's orbit can also expel gas from the Jovian potential well at a distance comparable to the truncation radius. Indeed, the removal of the circumplanetary nebula and the circumstellar disk must occur simultaneously.

%\\
%(start free-form ideas text) When gas removal happens, a relatively excited satellitesimal disk is retained. Maybe this disk forms the final satellites that don't migrate all the way in. In case of Jupiter, this is Callisto, in case of Saturn, it's Titan. Or maybe an instability is involved... who knows? This goes beyond the scope of our calculations, which only hope to outline the fundamentals of a new formation framework. Our calculations leave much to be desired in a detailed sense. Outline some future directions for work, and conclude...

\acknowledgments We are thankful to Katherine de Kleer, Darryl Seligman, Phil Hopkins, Mike Brown, and Christopher Spalding for insightful discussions. We thank Thomas Ronnet for providing a careful and insightful review of the manuscript. K.B. is grateful to the David and Lucile Packard Foundation and the Alfred P. Sloan Foundation for their generous support.

\begin{appendix} %\label{appnd}

\section{Dust Equilibrium from NSH Drift} \label{appndA}
%If needed

For simplicity, the analysis presented in section \ref{sec:dust} was carried out under the assumption that the cumulative back-reaction of solid dust upon gas is negligible. A more general description of dust-gas interactions within a nearly-Keplerian disk is provided by the Nakagawa--Sekiya--Hayashi drift. In a locally cartesian Keplerian ($\hat{x}=\hat{r},\hat{y}=r\,\hat{\phi}$) frame, the standard equations pertinent to the dust component of the system take the form \citep{1986Icar...67..375N}:
\begin{align}
&v_{r\,\bullet}=-\frac{2\,\eta\,\tau\,v_{\rm{K}}}{(1+\mathcal{Z})^2+\tau^2}\,\hat{x} \nonumber \\
&v_{\phi\,\bullet}=-\frac{\eta\,v_{\rm{K}}\,(1+\mathcal{Z})}{(1+\mathcal{Z})^2+\tau^2}+v_{\rm{K}}\,\hat{y}.
\label{vrvphi}
\end{align}
Qualitatively, the above expressions simply state that the primary outcome of gas-dust coupling is the inward drift of solids that is compensated by the outward expulsion of gas.

Importantly, these equations were derived assuming that the unperturbed azimuthal velocity of the gas is $v_{\phi}=(1-\eta)\,v_{\rm{K}}$ and that the unperturbed radial velocity is null. Conversely, if baseline radial velocity of the gas is $v_{r}$, by analogy with the second equation for the dust above, we have:
\begin{align}
&v_{r\,\bullet}=-\frac{2\,\eta\,\tau\,v_{\rm{K}}}{(1+\mathcal{Z})^2+\tau^2}+ \frac{v_r\,(1+\mathcal{Z})}{(1+\mathcal{Z})^2+\tau^2}\,\hat{x}.
\label{NSHmod}
\end{align}
Setting the RHS of this equation equal to zero, we obtain the equilibrium Stokes number that yields $v_{r\,\bullet}=0$:
\begin{align}
\tau^{(\rm{eq})}=\frac{v_{r}}{2\,\eta\,\vk}\big(1+\mathcal{Z}\big).
\label{StokeseqNSH}
\end{align}

\section{3D Bondi Accretion and the Transition to the Hill Regime} \label{appndB}
For convenience, in section \ref{sec:pebble}, we expressed the Bondi-Hill crossover mass, $\mathcal{M}_{\rm{t}}$ (equation \ref{eqn:mtrans}), as well as the rate of pebble accretion itself (equation \ref{eqn:BONDI3D}) in terms of global disk properties. Here we outline the derivation of these expressions.

The Bondi and Hill capture radii $\Rc$ and $\Rch$ are given by equations (\ref{eqn:Rc}) and (\ref{eqn:Rch}) respectively. As a starting step, we raise both quantities to the sixth power and set them equal to one another. Accordingly, $(\Rc)^6= (\Rch)^6$ gives
\begin{align}
\frac{25\,\mathcal{M}^2\,r^6\,v_r^2\,(1+\mathcal{Z}_{\rm{mid}})^2}{9\,\eta^2\,\M^2\,\vk^2}=\frac{\G^3\,\mathcal{M}^3\,r^3\,v_r^3\,(1+\mathcal{Z}_{\rm{mid}})^6}{8\,\eta^2\,\vk^9}.
\label{}
\end{align}
Rearranging for $\mathcal{M}$, we obtain
\begin{align}
\mathcal{M}=\frac{200\,\eta^4\,r^3\,\vk^7}{9\,\G^3\,\M^2\,v_r\,(1+\mathcal{Z}_{\rm{mid}})^4}.
\label{}
\end{align}
Substituting $2\,\pi\,r\,\Sigma/\Mdot$ for $1/v_r$ and multiplying the numerator as well as the denominator by $\M\,r$, yields
\begin{align}
\mathcal{M}=\frac{400\,\pi}{9}\,\M\,\frac{r^3}{\G^3\,\M^3}\, \frac{r^2\,\Sigma}{\Mdot}\,\vk^6\,\frac{\vk}{r}  \frac{\eta^4}{(1+\mathcal{Z}_{\rm{mid}})^4}.
\label{}
\end{align}
Finally, cancelling $(r/(\G\,\M))^3$ and $\vk^6$ while consolidating $\vk/r$ into $\Omega$, we arrive at equation (\ref{eqn:mtrans}):
\begin{align}
\mathcal{M}_{\rm{t}}=\frac{400\,\pi}{9}\,\bigg(\frac{\eta}{1+\mathcal{Z}_{\rm{mid}} } \bigg)^4\, \bigg( \frac{\Sigma\,r^2\,\Omega}{\Mdot} \bigg)\,\M.
\end{align}

The rate of accretion in the Bondi regime is derived in a similar manner. We begin by noting that the product of $(\Rc)^2$ and $\Delta v$ is
\begin{align}
\big(\Rc\big)^2\,\Delta v = \frac{\G\,\mathcal{M}\,r\,v_r\,(1+\mathcal{Z}_{\rm{mid}})}{2\,\eta\,\vk^2}.
\end{align}
Meanwhile, $\Sigma_{\bullet}=\mathcal{Z}\,\Sigma$ and $h_{\bullet}=r\sqrt{\alpha\,\epsilon/(2\,\mathcal{Z})} \,(h/r)$. Therefore,
\begin{align}
\frac{\Sigma_{\bullet}}{h_{\bullet}}=\frac{\Sigma}{r} \sqrt{\frac{2\,\mathcal{Z}^3}{\alpha\,\epsilon}}\bigg( \frac{h}{r} \bigg)^{-1}.
\end{align}
Multiplying the two expressions together while introducing a factor of $\M$ in both the numerator and the denominator, we have:  
\begin{align}
\big(\Rc\big)^2\,\Delta v\,\frac{\Sigma_{\bullet}}{h_{\bullet}}\,\frac{\M}{\M} = \frac{\mathcal{M}}{\M}\,\frac{\G\,\M}{r}\,\frac{r\,v_r\,\Sigma\,(1+\mathcal{Z}_{\rm{mid}})}{2\,\eta\,\vk^2}\bigg(\frac{h}{r}\bigg)^{-1}\, \sqrt{\frac{2\,\mathcal{Z}^3}{\alpha\,\epsilon}}.
\end{align}
Cancelling $\G\,\M/r$ with $1/\vk^2$ and multiplying the above expression by a factor of $\sqrt{\pi/2}$, we obtain equation (\ref{eqn:BONDI3D}):
\begin{align}
\bigg(\frac{d\,\mathcal{M}}{d\,t}\bigg)_{\rm{B \, 3D}} = \sqrt{\frac{\pi}{2}}\,\frac{\Sigma_{\bullet}}{h_{\bullet}}\,\big( \Rc \big)^2\,\Delta v =\frac{1}{2}\frac{\mathcal{M}}{\M}\frac{1+\mathcal{Z}_{\rm{mid}}}{\eta}\bigg( \frac{h}{r}\bigg)^{-1}\sqrt{\frac{\pi\,\mathcal{Z}^3}{\alpha\,\epsilon}}\,\Sigma\,r\,v_r.
\end{align}

\end{appendix}

\end{document}